\documentclass[sigconf, screen=true]{acmart}
\usepackage[utf8]{inputenc} % allow utf-8 input
\usepackage[T1]{fontenc}    % use 8-bit T1 fonts
\usepackage{hyperref}       % hyperlinks
\usepackage{amsfonts}       % blackboard math symbols
\usepackage{fancyhdr}       % header
\usepackage{graphicx}       % graphics

\usepackage{amsmath}
\usepackage{amsfonts}
\usepackage{amsthm}
\newtheorem{definition}{Definition}

\newtheorem{proposition}{Proposition}
\newtheorem{remark}{Remark}

\newtheorem{theorem}{Theorem}
\newcommand{\precdot}{\prec\mathrel{\mkern-5mu}\mathrel{\cdot}}

\makeatletter
\newcommand{\preceqdot}{\mathrel{\mathpalette\pr@ceqd@t\relax}}
\newcommand{\pr@ceqd@t}[2]{%
  \begingroup
  \sbox\z@{$#1\prec$}\sbox\tw@{$#1\preceq$}%
  \dimen@=\dimexpr\ht\tw@-\ht\z@\relax
  {\preceq}%
  \mkern-5mu
  \raisebox{\dimen@}{$\m@th#1\cdot$}%
  \endgroup
}
\makeatother

\newcommand{\ledot}{<\mathrel{\mkern-5mu}\mathrel{\cdot}}

\AtBeginDocument{%
  }

\copyrightyear{2022}
\acmYear{2022}
\setcopyright{acmlicensed}
\acmConference[ICTIR '22] {Proceedings of the 2022 ACM SIGIR International Conference on the Theory of Information Retrieval}{July 11--12, 2022}{Madrid, Spain.}
\acmPrice{15.00}
\acmISBN{978-1-4503-9412-3/22/07}
\acmDOI{10.1145/3539813.3545153}

\begin{document}

\title{On the Effect of Ranking Axioms on IR Evaluation Metrics}

\author{Fernando Giner}
\email{fginer3@gmail.com}
\orcid{1234-5678-9012}
\affiliation{%
  \institution{E.T.S.I. Informática UNED}
  \streetaddress{C/ Juan del Rosal, 16}
  \city{Madrid}
  \country{Spain}
  \postcode{28040}
}

\renewcommand{\shortauthors}{F. Giner}

\begin{abstract}
The study of IR evaluation metrics through axiomatic analysis enables a better understanding of their numerical properties. Some works have modelled the effectiveness of retrieval metrics with axioms that capture desirable properties on the set of rankings of documents. This paper formally explores the effect of these ranking axioms on the numerical values of some IR evaluation metrics. It focuses on the set of ranked lists of documents with multigrade relevance. The possible orderings in this set are derived from three commonly accepted ranking axioms on retrieval metrics; then, they are classified by their latticial properties. When relevant documents are prioritised, a subset of document rankings are identified: the \emph{join-irreducible} elements, which have some resemblance to the concept of basis in vector space. It is possible to compute the precision, recall, $RBP$ or $DCG$ values of any ranking from their values in the join-irreducible elements. However this is not the case when the swapping of documents is considered. 
\end{abstract}




\maketitle
\pagestyle{plain}
\pagenumbering{gobble}

\section{Introduction}
In the Cranfield paradigm \cite{cleverdon1967cranfield}, relevance judgements are assigned to a ranked list of documents for a topic. Then, an evaluation metric can be seen as a mapping that relates the set of possible lists of judged documents with a numeric structure, for instance, the set of real numbers. In this way, an evaluation metric assigns numbers (scores) that correspond to or 'preserve' certain observed relations between the set of possible lists of documents.

Many properties that an evaluation metric possesses are directly related to the properties of the set of rankings where it is defined, for example, in the statement ``an evaluation metric is an ordinal scale'', the evaluation metric is preserving the \emph{existing ordering} between rankings of documents. This characteristic has been observed in some axiomatic studies, such as \emph{linear algebra}, where a linear form, defined from a vector space to the real numbers, verifies that the image of any vector is determined by the image of a basis. It can also be seen in \emph{measurement theory}, where the representation theorems look for the requirements in the set where a measure is defined \cite{roberts1985measurement, krantz1971foundations, krantz1989foundations}. In these studies the set where the mapping is defined (empirical domain), endowed with relationships and\slash or operations, is distinguished from the numerical values that the mapping can reach (numerical image).

The empirical domain of the IR evaluation metrics can be shaped by desirable properties, for instance, the property ``it is preferred to retrieve relevant documents in the top ranking positions'' establishes a preference relation in the set of rankings of documents. The use of formal properties or axioms has been successfully applied to improve and understand evaluation metrics \cite{bollmann1984two,moffat2013seven,amigo2013general,sebastiani2015axiomatically,ferrante2015towards,amigo2018axiomatic, sebastiani2020evaluation}. The axiomatic approach of ``evaluating evaluation'' consists of describing desirable properties that a retrieval metric should satisfy, then the acceptability of a given metric is deduced or an evaluation metric can be axiomatically derived. We follow this axiomatic view and formally explore the effect of ranking axioms on the numerical properties of some IR evaluation metrics. %, modelling the set of ranked lists of documents so that any ranking can be expressed in terms of a distinguished subset of elements (the \emph{join-irreducible}). Then, the IR evaluation metrics that verify a kind of linearity, such as precision, recall, $RBP$ or $DCG$, are determined by their values on the subset of join-irreducible elements.

In this paper, intuitive and commonly accepted ranking axioms are considered, to ground the behaviour of IR evaluation metrics in a representative and well understood axiomatic basis. Then, these ranking axioms are operationalised and the possible orderings in the set of rankings of documents are derived. Applying results of lattice theory, a remarkable subset of elements is identified: the \emph{join-irreducible} elements. When relevant documents are prioritised, every ranked list of documents can be expressed in a unique manner as a join-combination of join-irreducible elements, similarly to the concept of \emph{basis} in vector space. In this context, IR evaluation metrics are real-mappings defined on the set of rankings. It is shown that precision, recall, $DCG$ and $RBP$ verify a kind of linearity, i.e., they are \emph{valuations} (in lattice theory terms). This property allows to determine their values on any ranking from the values on the join-irreducible elements; similarly to how the image of a linear form is determined by the images of the elements of a basis. However, when the swapping of documents is considered, they do not necessarily verify this property. In addition to understanding how IR evaluation metrics take effect on the set of output system runs, this result has some theoretical and practical consequences, especially in studies %that require large data collection 
devoted to test the behaviour of retrieval metrics.

This paper is organized as follows: In Section \ref{sec:SoA}, some related work is reported, and three operations on the set of ranked lists of documents are stated. In Section \ref{sec:lattice}, the basic notions of lattice theory are introduced. Section \ref{sec:notation} formalizes the problem in the context of IR. Sections \ref{sec:set-based} and \ref{sec:rank-based} contain the main results, where the possible orderings on the set of rankings of documents and their structural properties are derived for the set-based and rank-based retrieval. Finally, in Section \ref{sec:conclusion} some conclusions and applications are drawn.

\section{Related Work}
\label{sec:SoA} 
The formal analysis of IR evaluation metrics has contributed to a better understanding of their properties. One of the first attempts is given in \cite{swets1963information}, where the effectiveness and the efficiency of IR evaluation metrics are analized in terms of a $2$-by-$2$ contingency table of pertinence and retrieval. Later, van Rijsbergen \cite{van1974foundation,van1979information}, tackle the issue of the foundations of measurement in IR through a conjoint (additive) structure based on precision and recall; then, he examines the properties of a measure on this prec-recall structure. In \cite{bollmann1980measurement}, a similar conjoint structure is defined, but on the contingency table of the binary retrieval; then, they study the properties of the proposed MZ-metric. In \cite{bollmann1984two}, two axioms are highlighted, then, all the IR metrics, which are compliant with them, are expressed as a linear combination of the number of relevant\slash non-relevant retrieved documents. In \cite{yao1995measuring}, user judgements on documents are formally described by a weak order, then, a measure of system performance is presented, whose appropriateness is
demonstrated through an axiomatic approach. In \cite{huibers1996axiomatic}, a framework for the theoretical comparison of IR models, based on situation theory, is presented. It enables an inference mechanism with the axiomatised concept of \emph{aboutness}. In \cite{amigo2013general,amigo2009comparison}, some properties of evaluation metrics for clustering are formalized, describing how IR metrics should behave. In \cite{moffat2013seven}, it is proposed a formal framework based on numerical properties of IR evaluation metrics, it helps to better understand the relative merits of IR metrics for different applications. In \cite{sebastiani2015axiomatically}, the axioms that an evaluation measure for classification should satisfy are discussed, then the K evaluation metric is proposed. In \cite{busin2013axiometrics, maddalena2014axiometrics} an axiomatic definition of IR effectiveness metric is provided, in a unifying framework for ranking, classification and clustering. In \cite{sebastiani2020evaluation}, it is discussed what properties should enjoy an evaluation measure for quantification. 

In recent years, some forums and conferences \cite{zhai2013axiomatic, amigo2017axiomatic,amigo2020axiomatic, maddalena2014axiometrics, busin2013axiometrics} have encouraged the search for desirable properties expressed mathematically as formal constraints to assess the optimality of retrieval models. They are formal properties that a ``good'' retrieval model should fulfil \cite{fang2011diagnostic,fang2004formal, fang2006semantic,fang2005exploration}. This analytical approach allows to explore retrieval models and how best to improve them, in order to achieve higher retrieval effectiveness. It has been successfully applied to the study of basic models \cite{fang2004formal, fang2007axiomatic}, pseudo-relevance feedback methods \cite{clinchant2011document,clinchant2013theoretical, montazeralghaem2016axiomatic}, translation retrieval models \cite{karimzadehgan2012axiomatic,rahimi2020axiomatic} and neural network retrieval models \cite{rosset2019axiomatic}.

In \cite{amigo2020nature}, a survey about the attempts to formalize the properties of evaluation metrics is provided. They consider a wide variety of information access tasks and classify these properties into four categories: (i) general axioms \cite{moffat2013seven,sebastiani2015axiomatically}, (ii) axioms for classification metrics \cite{sokolova2006assessing, sebastiani2015axiomatically, amigo2017axiomatic2}, (iii) axioms for clustering metrics \cite{dom2012information,rosenberg2007v, amigo2009comparison, meilua2007comparing} and (iv) ranking axioms \cite{bollmann1984two,moffat2013seven,ferrante2015towards,amigo2013general,amigo2018axiomatic}.

\subsection{Ranking Axioms}
In terms of the batch evaluation of IR systems, some works have highlighted formal properties in the set of rankings. In \cite{bollmann1984two}, two axioms are defined: the \emph{monotonicity} and the \emph{Archimedian}, which determine the form of the evaluation metrics from a measurement theoretic perspective. In  \cite{moffat2013seven}, the evaluation metrics are characterized by seven numeric properties, such as \emph{convergence}, which states that relevant documents must occur above non-relevant ones, \emph{top-weightedness}, \emph{localization} and others. It provides a framework to classify the metrics according to their effectiveness. In \cite{amigo2013general, amigo2018axiomatic}, the following properties are highlighted: \emph{priority}, which states that swapping documents in a correct way is a preferred option, \emph{deepness}, which affirm that deeper  positions  in the ranking are less likely to be explored, \emph{closeness threshold}, which states that there is an area at the top of the ranking which is always explored, \emph{confidence}, etc. They derive evaluation metrics, which can be applied to several information access tasks. In \cite{ferrante2018general}, the set of rankings of documents is characterized by the \emph{replacement} operation, which prioritize the replacement of a document by another document with higher relevance degree; and the \emph{swap} operation, which states that swapping a less relevant document with a more relevant one in a lower rank position is a preferred option. In \cite{amigo2020nature}, the swapping of documents is highlighted since it is common to many other tasks.

We conduct our analysis on the basis of three formal properties, which express preferences on the empirical domain of the retrieval metrics. They are explicitly or implicitly present in most of these works: (i) \emph{relevance}: ``a run retrieving relevant documents is preferred to another one retrieving less relevant documents'', this property has been studied under several names such as \emph{monotonicity} \cite{bollmann1984two}, \emph{replacement} \cite{ferrante2015towards} or \emph{confidence} \cite{amigo2018axiomatic}; (ii) \emph{deepness}: ``it is better one relevant document in the highest part of a list than many other relevant documents in the bottom part'', this property is stated in \cite{amigo2013general,amigo2018axiomatic} as a \emph{deepness threshold} or \emph{deepness} and they assert that one relevant document in the highest part of a list is preferred over many other relevant documents in the bottom part; and (iii) \emph{swapping}: ``swapping a less relevant document in a higher rank position with a more relevant one in a lower rank position is a preferred option'', this property is known as \emph{convergence} \cite{moffat2013seven}, \emph{swap} \cite{ferrante2015towards,amigo2020nature} or \emph{priority constraint} \cite{amigo2013general}.

These ranking axioms can be expressed in terms of ranking operations \cite{hagen2016axiomatic, volske2021towards}, it enables to explore to what extent the numerical properties of IR evaluation metrics can be explained by the effect of ranking axioms, and which axioms apply in these situations. This characteristic has been shown in \cite{ferrante2018general}, where the orderings between rankings are based on two operations: \emph{replacement} and \emph{swap}. In this paper, the ranking constraints are operationalized as Boolean predicates which, given a pair of ranked lists of documents, express a preference between them. 

\begin{quote}
\textbf{Operation 1 (Replacement):}

``Replacing a document for another one with a higher relevance degree is a preferred option''.
\end{quote}
This operation implements the \emph{relevance} axiom and it establishes a simple property: \emph{prioritizing relevant documents}.
 
\begin{quote}
\textbf{Operation 2 (Projection):}

``Removing all documents that are not considered in the highest or more relevant part of a list is a preferred option''.
\end{quote}
This operation implements the \emph{deepness} axiom and it remarks on a specific part (the highest or more notable) where it should be considered a relevant document, without paying attention to the rest of the documents. 

\begin{quote}
\textbf{Operation 3 (Swapping):}

``Swapping a less relevant document in a higher rank position with a more relevant one in a lower rank position is a preferred option''.
\end{quote}
This operation implements the \emph{swapping} axiom.

These three operations express preferences between ranked lists of documents, thus, every operation give rise to a total or partial ordering in the empirical domain of IR evaluation metrics. One of the major advantages of having operationalized the axioms is that it facilitates the combinations of them. In Sections \ref{sec:set-based} and \ref{sec:rank-based}, the possible combinations of these operations are considered, resulting in five different orderings in the set of ranked lists of documents. 

\section{Lattice Theory}
\label{sec:lattice}
In this section we recall some basic notions of lattice theory, for additional background, we refer the reader to the textbooks \cite{birkhoff1940lattice, gratzer2002general}. 

A \emph{partially ordered set} or \emph{poset} is a non-empty set, $X$, together with a binary, reflexive, antisymmetric and transitive relation, $\preceq$, defined on $X$. A poset $(X, \preceq)$ is called a \emph{totally ordered set} or a \emph{chain} if any two elements, $x$, $y \in X$, are comparable, i.e., $x \preceq y$ or $y \preceq x$.

A (closed) interval is a subset of $X$ defined as $[x, y] = \{z \in X : x \preceq z \preceq y\}$. It is said that $y$ \emph{covers} $x$ in $X$, denoted by $x \precdot y$, if there does not exist $z \in X$ such that $x \neq z \neq y$ and $x \preceq z \preceq y$\footnote{A binary relation of order can be described in two ways. For example, the relation ``two real numbers are related if one of them is, at least, one higher unity than the other'', can be described: (1) in an extensive way: $a \leq b \Leftrightarrow a \leq b + 1$, $\forall a, b \in \mathbb{R}$, i.e, describing all the possible pairs which are related; or (2) with the covering relation: $a \ledot b \Leftrightarrow a = b + 1$, $\forall a, b \in \mathbb{R}$, i.e. describing only the pairs which one element covers the other.}.

Thus, $(X, \preceq)$ has an associated graph, called the \emph{Hasse diagram}, where nodes are labelled with the elements of $X$ and the edges indicate the covering relation. Nodes are represented in different levels, if $y$ covers $x$, then $x$ is below $y$ in the diagram. 

A noteworthy property of a finite poset is that its covering relation is the transitive reduction of the partial order relation, i.e., the covering relation cannot contain the following pairs $x \precdot y$, $y \precdot z$ and $x \precdot z$ at the same time, for any $x$, $y$, $z \in X$.

An upper bound for a subset, $S \subseteq X$, is an element, $x \in X$, for which $s \preceq x$, $\forall s \in S$. A lower bound for a subset of a poset is defined analogously. The greatest lower bound or \emph{meet} of two elements $x, y \in X$ is denoted by $x \wedge y$. The least upper bound or \emph{join} of two elements $x, y \in X$ is denoted by $x \vee y$. It is said that $X$ has a \emph{top element}, denoted by $1 \in X$, if $x \preceq 1$, $\forall x \in X$; dually, $0 \in X$ is the \emph{bottom element}, if $0 \preceq x$, $\forall x \in X$.

Considering the join and the meet as operators, $(X, \wedge, \vee, \preceq)$ is a \emph{lattice} if $(X, \preceq)$ is a poset and every two elements of $X$ have a meet and a join. A \emph{sublattice}, $S \subseteq X$, is a subset of the lattice that is closed to the meet and join operations, i.e., $x\wedge y$, $x \vee y \in X$, $\forall x$, $y \in S$.

A lattice is \emph{distributive} if its meet operator distributes over its join operator, i.e., $x \wedge (y \vee z)$ $=$ $(x \wedge y) \vee (x \wedge y)$, $\forall x$, $y$, $z \in X$. 

\begin{figure}[h]
  \centering
  \scalebox{0.7}
  {
  \includegraphics[width=\linewidth]{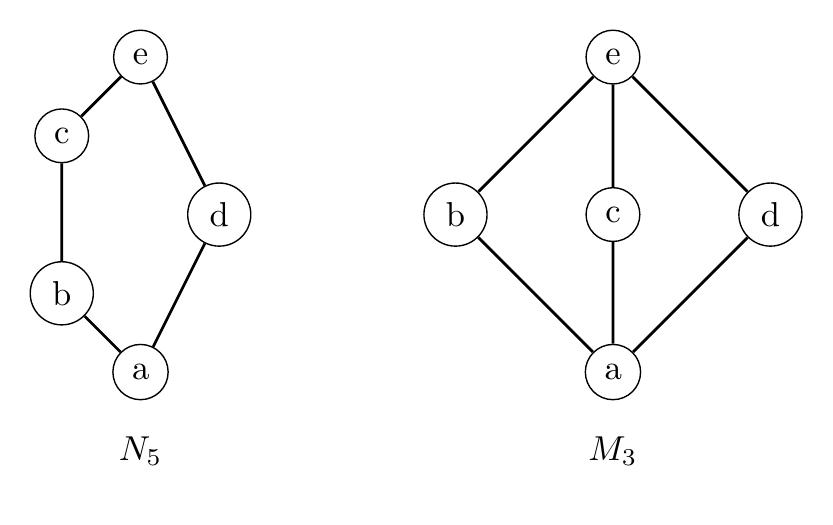}
  }
  \caption{Lattices $N_5$ and $M_3$ are not distributive}
	\label{fig:distributive}
  \Description{The distributive lattices $N_5$ and $M_3$.}
\end{figure}

For instance, any chain is a distributive lattice \cite{birkhoff1940lattice}. There are two specific lattices that characterize the distributive lattices, the diamond lattice ($M_3$) and the pentagon lattice ($N_5$), which are shown in Figure \ref{fig:distributive}.
\begin{theorem}[Birkhoff \cite{birkhoff1940lattice}] \label{th:distributive_n_m}
A lattice, $X$, is distributive, if and only if it does not contain $M_3$ or $N_5$ as sublattices.
\end{theorem}

There are some remarkable elements in a lattice, given $j \in X$ (with $j \neq 0$), it is said to be a \emph{join-irreducible} element, if $x \vee y = j$ implies $x = j$ or $y = j$. Thus, $j \in X$ is join-irreducible if it cannot be expressed as the join of two elements that are strictly smaller than $j$. We denote the set of join-irreducible elements of $X$ by $J_{X}$\footnote{The join-irreducible elements are those that have only one descendant in the covering graph \cite{gratzer2002general}}. The following result highlights the importance of the join-irreducible elements of a finite distributive lattice.
\begin{theorem}[Blyth \cite{blyth2005distributive}] \label{th:expresion-join}
If $X$ is a finite distributive lattice, then every element of $X \setminus \{0\}$ can be expressed uniquely as an irredundant join of join-irreducible elements.
\end{theorem}
Join-irreducible elements have some resemblance to the concept of basis in vector space since every element is determined by them. 

A real-valued mapping defined on a lattice, $v: X \rightarrow \mathbb{R}$, is a \emph{valuation}, if it satisfies the following property: $v(x) + v(y) = v(x \vee y) + v(x \wedge y)$, $\forall x$, $y \in X$. 

The next result shows a characterization of the valuations in terms of the join-irreducible elements.
\begin{theorem}[Rota \cite{rota1971combinatorics}] \label{th:valuation-distributive}
A valuation, defined on a finite distributive lattice $X$, is uniquely determined by the values it takes on the set of join-irreducible elements of $X$. 
\end{theorem}

\section{IR Setting}
\label{sec:notation}
In this section, some notation needed throughout the paper is introduced. The formalism of \cite{ferrante2018general} is adopted.

Consider a finite set of documents that are retrieved from query terms provided by a user. An (ordered or not) collection of $N$ retrieved documents is called the \emph{system run} of length $N$ for a topic. 

Once documents have been retrieved, they can be classified by their relevance degree. Let $(REL, \preccurlyeq)$ be a finite and totally ordered set of relevance degrees, where $REL = \{\mathit{a}_0, \ldots, \mathit{a}_{c}\}$ with $\mathit{a}_{i} \prec \mathit{a}_{i+1}$, for each $i \in \{1, \ldots, c-1\}$. These relevance degrees can be categorical labels. To handle numerical values, a \emph{gain function}, $g: REL \rightarrow \mathbb{R}^{+}$, is considered as a map that assigns a positive real number to any relevance degree. In this paper, it can be assumed that $g(\mathit{a}_0) = 0$ and $g$ is a strictly increasing function. 

When every retrieved document of a system run is classified with a relevance degree, a \emph{judged run} is obtained, denoted by $\hat{r}$. 

In \emph{set-based retrieval}, each judged run is an unordered set of $N$ relevance degrees, which may contain the same element several times. For instance, for $N=4$ and $3$ relevance degrees, a judged run is given by $\hat{r} = \{\mathit{a}_2, \mathit{a}_1, \mathit{a}_1, \mathit{a}_0\}$. For the sake of clarity, the convention is used to represent $\hat{r}$ as $\{\hat{r}_1, \ldots, \hat{r}_N\}$, where $\hat{r}_i \succcurlyeq \hat{r}_{i + 1}$, for any $i \in \{1, \ldots N-1\}$, i.e., the relevance degrees are listed in decreasing order. This process does not affect the results obtained. The $j$-th element of the set $\hat{r}$ is denoted by $\hat{r}_j$.

In \emph{rank-based retrieval}, each judged run is an ordered list of $N$ relevance degrees. For instance, for $N=4$ and $3$ relevance degrees, a judged run is given by $\hat{r} = (\mathit{a}_1, \mathit{a}_0, \mathit{a}_2, \mathit{a}_1)$. In this case, the $j$-th element of $\hat{r}$ is denoted by $\hat{r}[j]$ .
 
The set of all the possible judged runs of length $N$ is a finite set, denoted by $R(N)$.

The \emph{replacement} operation of Section \ref{sec:SoA} establishes that relevant documents are prioritized. This property can be quantified with the following concept. 
\begin{definition}
The \emph{cumulated mass of relevance} of a judged run, $\hat{r} \in R(N)$, for a relevance degree, $\mathit{a}_j \in REL$, is the number of documents of $\hat{r}$ with a higher relevance degree than $\mathit{a}_j$.

In set-based retrieval, considering that $\hat{r} =\{ \hat{r}_1, \ldots, \hat{r}_N\} \in R(N)$, it is given by $\big\vert \{ i : \hat{r}_i \succcurlyeq \mathit{a}_j \} \big\vert$. In rank-based retrieval, considering that $\hat{r} =(\hat{r}[1], \ldots, \hat{r}[N]) \in R(N)$, it is given by $\big\vert \{ i : \hat{r}[i] \succcurlyeq \mathit{a}_j \} \big\vert$
\end{definition}
\begin{remark}
In rank-based retrieval, this definition can be generalized to a specific position of the ranking. The \emph{cumulated mass of relevance}, for $\mathit{a}_j \in REL$, at position $k$, where $1 \leq k \leq N$, is the number of documents with a higher relevance degree than $\mathit{a}_j$, from the first to the $k$-th position, which is given by $\big\vert \{1 \leq i \leq k : \hat{r}[i] \succcurlyeq \mathit{a}_j \} \big\vert$
\end{remark}

The \emph{projection} operation establishes that the documents that are not considered in the highest or most relevant part of the list should be removed. In rank-based retrieval, the highest or most relevant part of the list is specified by the top positions of the ranking. Thus, this property can be understood as removing those documents below a specific position in the ranking. On the other hand, in the set-based retrieval, there are no positions of a ranking, but instead the more relevant part of the list is specified by the relevance degrees. Thus, this operation can be understood as removing those documents that do not have high relevance degrees. 

Considering different combinations of the three properties of Section \ref{sec:SoA}, preference relations will be obtained on the set of judged runs, i.e., $R(N)$ will be endowed with an ordering, which will be denoted by $\preceq$. 

If this ordering is a poset, then $(R(N), \wedge, \vee, \preceq)$ can be considered as a lattice, since $R(N)$ is a finite set, where every pair of elements have a meet and join, trivially.

An evaluation metric can be seen as a mapping between $R(N)$ and the set of real numbers. In set based retrieval some remarkable evaluation measures are as follows.
\begin{itemize}
\item Generalized precision ($gP$) \cite{ferrante2018general} :
$$gP(\hat{r}) = \frac{1}{N} \cdot \sum_{i=1}^{N} \frac{g(\hat{r_i})}{g(\mathit{a}_{c})} \ .$$
\item Generalized recall ($gR$) \cite{ferrante2018general}
$$gR(\hat{r}) = \frac{1}{RB} \cdot \sum_{i=1}^{N} \frac{g(\hat{r_i})}{g(\mathit{a}_{c})} \ ,$$
where $RB$ is the recall base (total number of relevant documents).
\end{itemize}
These evaluation measures can also be considered in rank-based retrieval, in addition to the following.
\begin{itemize}
\item Graded rank-biased precision ($gRBP$) \cite{moffat2008rank,sakai2008information, ferrante2018general}:
$$gRBP(\hat{r}) = \frac{(1-p)}{g(\mathit{a}_c)} \cdot \sum_{i=1}^N p^{i-1} \cdot g(\hat{r}[i]) \ .$$
\item Discounted cumulated gain ($DCG$) \cite{kekalainen2002using}:
$$DCG_{b}(\hat{r}) = \sum_{i=1}^N \frac{g(\hat{r[i]})}{\ \max \{1, \log_{b}i \}} \ ,$$
\end{itemize}

%The operations presented in Section \ref{sec:SoA}, express pairwise preferences for judged runs, $\hat{r}$ and $\hat{s}$, thus, every operation induces an ordering with a preference, $\hat{r} \preceq \hat{s}$, which can be represented by its Hasse diagram (see Section \ref{sec:lattice}). 
In the two following sections, the possible orderings in set-based and rank-based retrieval are deduced by means of possible combinations of the three operations presented (see Section \ref{sec:SoA}). It has to be noted that considering multiple operations at the same time, could yield a set of possibly conflicting preferences for ranked lists of documents. In this case, those combinations of operations that generate an ordering relation with cycles are dismissed since they do not satisfy the transitivity of a partial order. Likewise, those combinations of axioms that generate a disconnected Hasse diagram, i.e., the cases where there exists an element or a group of elements that cannot be compared with the rest of the elements, are also excluded. In these cases, the obtained results can be applied to each connected component of the Hasse diagram.

\section{Set-based Retrieval}
\label{sec:set-based}
In this scenario, a system run is a set of relevance degrees, where the judged documents are listed in decreasing order.

It is not possible to consider the \emph{swapping} operation since interchanging documents in a set has no effect. On the other hand, the \emph{replacement} operation prioritizes relevant documents, and the \emph{projection} operation focuses on the most relevant part of the list. 

Thus, the possible combinations of operations will be: (i) considering the \emph{projection} and the \emph{replacement} operations jointly  and (ii) considering the \emph{replacement} operation individually.

Note that it is not possible to consider the \emph{projection} operation individually since it refers to a specific part of the list, and the \emph{replacement} operation is needed to prioritize relevant documents.

\subsection{Projection+Replacement [set-based]}
Given a pair of judged runs, the \emph{projection} operation selects the more relevant part of the lists, i.e., the highest relevance degree at which the two system runs differ. In addition, considering the \emph{replacement} operation, the ranking that has greater \emph{cumulative mass of relevance} (or equivalently more relevant  documents\footnote{Considering the highest relevance degree at which the two system runs differ, the difference between the \emph{cumulated mass of relevance} is determined by the number of documents with that relevance degree.}) for that relevance degree is preferred. Formally:
\begin{quote}
\textbf{Projection+Replacement} [set-based]

Let $\hat{r}, \hat{s} \in R(N)$ such that $\hat{r} \neq \hat{s}$, and let $k$ be the largest relevance degree at which the two runs differ for the first time, i.e., 
$k = \max \big\{j \leq c : \big\vert \{i : \hat{r}_i = \mathit{a}_j\}\big\vert \neq \big\vert \{i : \hat{s}_i = \mathit{a}_j\} \big\vert \big\}$, then they are ordered by: 
$$\hat{r} \preceq \hat{s} \Longleftrightarrow \big\vert \{i : \hat{r}_i = \mathit{a}_k\}\big\vert \leq \big\vert \{i : \hat{s}_i = \mathit{a}_k\}\big\vert$$
\end{quote}
A system run is preferred to another, if for the highest relevance degree at which the two systems runs differ, the first has more documents with this relevance degree.

$(R(N, \preceq)$ is a totally ordered set or a chain\footnote{This ordering is the same as the ``Total Ordering'' (set-based case) studied in \cite{ferrante2018general}, where it can be seen the linearity of the order.}. Thus, the meet and join operations of a pair of elements are the minimum and maximum of both, respectively, i.e., $\hat{r} \wedge \hat{s} = \min \{ \hat{r}, \hat{s} \}$, $\hat{r} \vee \hat{s} = \max \{ \hat{r}, \hat{s} \}$.

As noted in Section \ref{sec:lattice}, $R(N)$ endowed with these operations is a distributive lattice since it is a chain. Moreover, every element of $R(N)$ is join-irreducible, except $\hat{0}$, since these elements have only one descendant in the Hasse diagram, $J_{R(N)} = R(N) \setminus \{\hat{0}\}$. Then, it is possible to apply Theorem \ref{th:valuation-distributive}, which states that the values reached by an evaluation metric are determined by the values of the join-irreducible elements. However, this is a trivial case where no relevant information is obtained since the values reached by an evaluation measure are determined by every ranking of the empirical domain, except $\hat{0}$.

\subsection{Replacement [set-based]}
Considering the \emph{replacement} operation individually, the relevant documents have to be prioritized, i.e., a system run is preferred to another, if it has more \emph{cumulated mass of relevance} for any relevance degree. 
Formally:
\begin{quote}
\textbf{Replacement} [set-based]

Any pair of system runs, $\hat{r}, \hat{s} \in R(N)$, such that $\hat{r} \neq \hat{s}$, is ordered by:
$$\hat{r} \preceq \hat{s} \Longleftrightarrow \big\vert\{i : \hat{r}_i \succcurlyeq \mathit{a}_j\}\big\vert \leq \big\vert\{i : \hat{s}_i \succcurlyeq \mathit{a}_j\}\big\vert, \forall j \in \{0, \ldots, c\}$$
\end{quote}
This ordering considers a run greater than another if, fixing any relevance degree, it has more documents above that relevance degree than does the other. 

In this ordering not every pair of rankings is comparable, for instance, $\hat{r} = \{\mathit{a}_2, \mathit{a}_2, \mathit{a}_0\}$ and $\hat{s} = \{\mathit{a}_3, \mathit{a}_0, \mathit{a}_0\}$, since $\big\vert\{i : \hat{r}_i \succcurlyeq \mathit{a}_3\}\big\vert = 0 < 1 = \big\vert\{i : \hat{s}_i \succcurlyeq \mathit{a}_3\}\big\vert$, while $\big\vert\{i : \hat{r}_i \succcurlyeq \mathit{a}_2\}\big\vert = 2 > 1 = \big\vert\{i : \hat{s}_i \succcurlyeq \mathit{a}_2\}\big\vert$.

\begin{figure}[bt]
  \centering
  \scalebox{0.35}
  {
  \includegraphics[width=\linewidth]{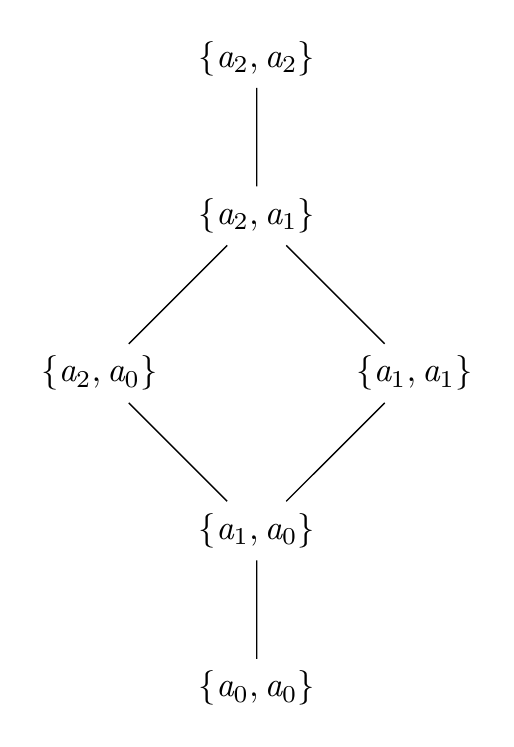}
  }
  \caption{Hasse diagram for $c=2$ and $N=2$ in the set-based retrieval}
	\label{fig:ini}
  \Description{Hasse diagram of the example ($c=2$ and $N=2$) in the set-based retrieval.}
\end{figure}

However, $R(N)$ endowed with this ordering is a \emph{poset}\footnote{This partial order is the same as the ``Partial Ordering'' (set-based case) studied in \cite{ferrante2018general}, where it can be seen this result.}. Figure \ref{fig:ini} shows the Hasse diagram for the case $c=2$ and $N=2$.

Considering the convention to represent judged runs in decreasing order, the join and the meet operations have an explicit expression. 
\begin{proposition}
\label{prop:operations-set}
Given $\hat{r} = \{\hat{r}_1, \ldots, \hat{r}_N\}$ and $\hat{s} = \{\hat{s}_1, \ldots, \hat{s}_N\}$ in $(R(N), \wedge, \vee, \preceq)$, with the \textbf{replacement} [set-based] partial order, their meet and join can be expressed as:
\begin{align*}
\hat{r} \wedge \hat{s} & = \Big\{ \min \{ \hat{r}_1 , \hat{s}_1 \}, \ldots, \min \{ \hat{r}_N , \hat{s}_{N} \} \Big\} \\
\hat{r} \vee \hat{s} & = \Big\{ \max \{ \hat{r}_1 , \hat{s}_1 \}, \ldots, \max \{ \hat{r}_N , \hat{s}_N \} \Big\}
\end{align*}
\end{proposition}
In this case, $(R(N), \preceq)$, endowed with the meet and join operations, is trivially a lattice, since $R(N)$ is a finite set. The following result shows the structure of $R(N)$.
\begin{proposition}
\label{prop:distrib-set}
$(R(N), \wedge, \vee, \preceq)$ is a distributive lattice, where $\preceq$ is the \textbf{replacement} [set-based] partial order.
\end{proposition}

In this context, a remarkable subset of judged runs is identified by the following result.
\begin{proposition}
\label{prop:irred-set}
The join-irreducible elements of $(R(N), \wedge, \vee, \preceq)$, where $\preceq$ is the \textbf{replacement} [set-based] ordering, are:
\begin{small}
\begin{equation*}
\begin{matrix}
\{\mathit{a}_1, \mathit{a}_0, \ldots, \mathit{a}_0 \}, & \{\mathit{a}_1, \mathit{a}_1, \mathit{a}_0, \ldots, \mathit{a}_0 \}, & \ldots, & \{\mathit{a}_1, \ldots, \mathit{a}_1 \},  \\
\{\mathit{a}_2, \mathit{a}_0, \ldots, \mathit{a}_0 \}, & \{\mathit{a}_2, \mathit{a}_2, \mathit{a}_0, \ldots, \mathit{a}_0 \}, & \ldots, & \{\mathit{a}_2, \ldots, \mathit{a}_2 \},  \\
\vdots & \vdots & \vdots & \vdots  \\
\{\mathit{a}_c, \mathit{a}_0, \ldots, \mathit{a}_0 \}, & \{\mathit{a}_c, \mathit{a}_c, \mathit{a}_0, \ldots, \mathit{a}_0 \}, & \ldots, & \{\mathit{a}_c, \ldots, \mathit{a}_c \} 
\end{matrix}
\end{equation*}
\end{small} 
\end{proposition}

\begin{figure}[bt]
  \centering
  \scalebox{1.06}
  {
  \includegraphics[width=\linewidth]{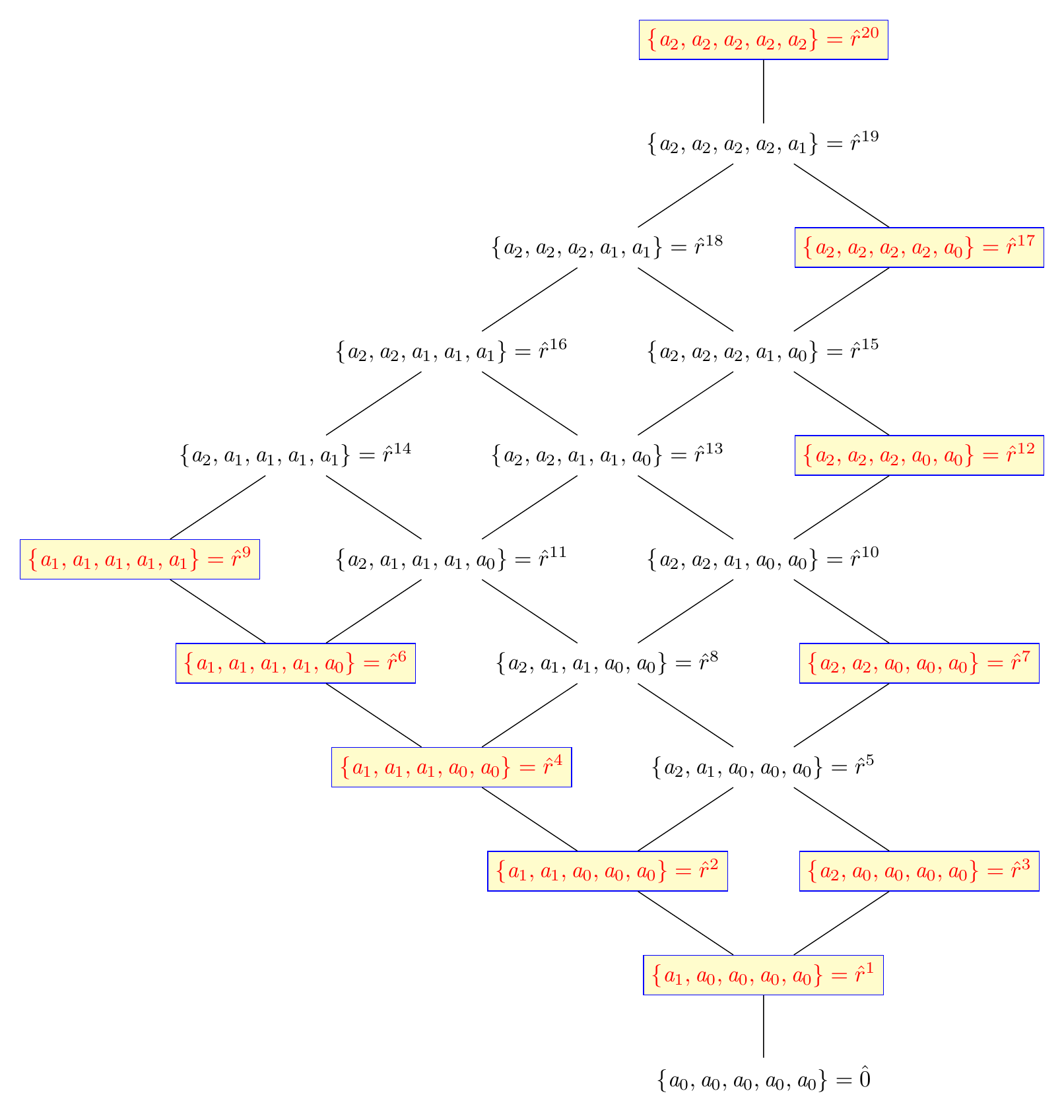}
  }
  \caption{Hasse diagram for $c=2$ and $N=5$ in the set-based retrieval}
	\label{fig:N-4-joinirreducibles2}
  \Description{Hasse diagram of the example ($c=2$ and $N=5$) in the set-based retrieval.}
\end{figure}

For example, for the case $N=5$ and three relevance degrees ($c=2$), the Hasse diagram is shown in Figure \ref{fig:N-4-joinirreducibles2}, where the join-irreducible elements have been marked.

Proposition \ref{prop:operations-set} provides an analytical way to determine the meet and the join of two judged runs. A graphical way to obtain the same results is illustrated in Figure \ref{fig:N-4-joinirreducibles2}, where the join of two elements is the intersection of the minimum upwards paths starting from these elements, and the meet of two judged runs is the intersection of the minimum downwards paths.

The result obtained in Proposition \ref{prop:distrib-set} allows us to apply Theorem \ref{th:expresion-join} to this partial order, showing the importance of the join-irreducible elements. It states that every judged run can be expressed as a join of the join-irreducible elements. This result can be checked for some judged runs of this example:
\begin{equation*}
\begin{matrix}
\hat{r}^5 = \hat{r}^2 \vee \hat{r}^3,  & \hat{r}^8 = \hat{r}^3 \vee \hat{r}^4,  & \hat{r}^{10} = \hat{r}^4 \vee \hat{r}^7,  & \hat{r}^{11} = \hat{r}^3 \vee \hat{r}^6,  \\
\hat{r}^{14} = \hat{r}^3 \vee \hat{r}^9,  & \hat{r}^{15} = \hat{r}^6 \vee \hat{r}^{12},  & \hat{r}^{16} = \hat{r}^7 \vee \hat{r}^9, &\hat{r}^{18} = \hat{r}^9 \vee \hat{r}^{12} 
\end{matrix}
\end{equation*}

\begin{figure}[bt]
  \centering
  %\scalebox{0.95}
  %{
  \includegraphics[width=\linewidth]{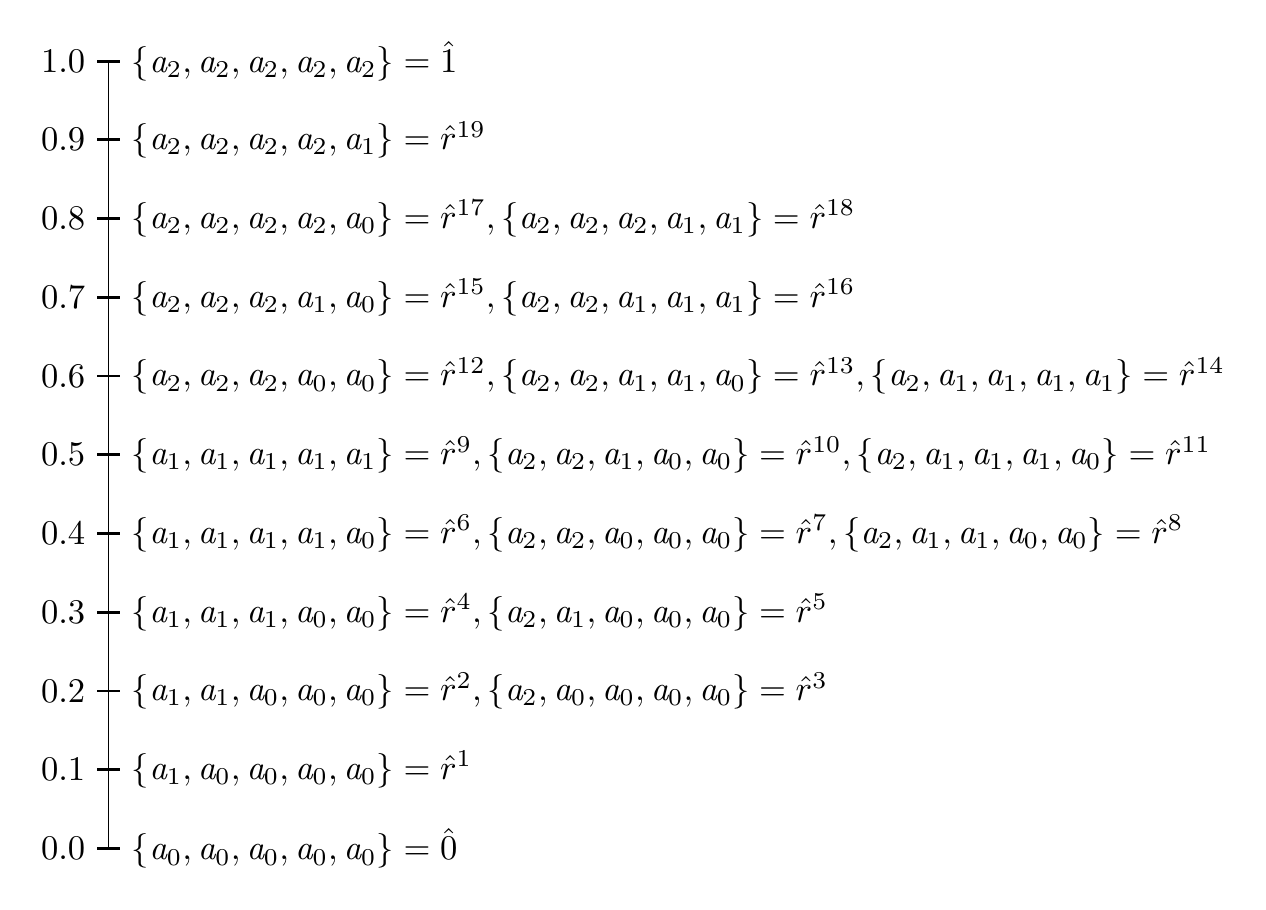}
  %}
  \caption{Generalized Precision and Recall for $c=2$ and $N=5$ in the set-based retrieval}
	\label{fig:prec-recall-set}
  \Description{Values of the generalized Precision and Recall ($c=2$ and $N=5$) in the set-based retrieval.}
\end{figure}

Once the properties of the empirical domain have been analysed, it is possible to study the properties of the evaluation measures defined on it. Figure \ref{fig:prec-recall-set} shows the numerical values reached by the generalized precision and recall (they are the same) for this example. These values have been obtained with the definition of $gP$ in Section \ref{sec:notation}.

In Section \ref{sec:lattice}, it has been shown that if a real mapping verifies a kind of linearity, i.e., if it is a valuation, then it is completely determined by the values of the join-irreducible elements. The following result shows that the IR evaluation metrics of this scenario verify this property.
\begin{proposition}
\label{prop:valuation-set}
The generalized precision and recall are valuations, when they are defined on $(R(N), \wedge, \vee, \preceq)$, where $\preceq$ is the \textbf{relevance} [set-based] partial order.
\end{proposition}

The main result of this subsection is obtained from Propositions \ref{prop:distrib-set} and \ref{prop:valuation-set}, which allow to apply Theorem \ref{th:valuation-distributive} to this partial order. Thus, the values reached by $gP$ and $gR$ are determined by the values reached on the set of join-irreducible elements.

Let us now look at some practical aspects of calculating these values. By means of Proposition \ref{prop:valuation-set}, the numerical values of $gP$ and $gR$ can be computed with the characterization property of valuations: $v(\hat{r} \vee \hat{s}) = v(\hat{r}) + v(\hat{s}) - v(\hat{r} \wedge \hat{s})$. Note that the last operand of the second part of the equality contains a meet of two join-irreducible elements. Considering Propositions \ref{prop:operations-set} and \ref{prop:irred-set}, it can be easily seen that every meet of any pair of join-irreducible elements is a join-irreducible. Thus, the characterization of valuations is sufficient to determine any numerical value. 

For example, Figure \ref{fig:prec-recall-set} shows that $gP(\hat{r}^5) = 0.3$. This value can also be obtained from the values of $gP$ in the set of join-irreducible elements. Making use of Theorem \ref{th:expresion-join} and Proposition \ref{prop:valuation-set}, $gP(\hat{r}^5) = gP(\hat{r}^2 \vee \hat{r}^3)  = gP(\hat{r}^2) + gP(\hat{r}^3) - gP(\hat{r}^2 \wedge \hat{r}^3) = 0.2 + 0.2 - gP(\hat{r}^1) = 0.2 + 0.2 - 0.1 = 0.3$.

The following equalities check some numerical values of the $gP$ evaluation measure.
\begin{align*}
& gP(\hat{r}^5) = 0.2 + 0.2 - 0.1 = 0.3 & gP(\hat{r}^8) = 0.2 + 0.3 - 0.1 = 0.4 & \\
& gP(\hat{r}^{10}) = 0.3 + 0.4 - 0.2 = 0.5 & gP(\hat{r}^{11}) = 0.2 + 0.4 - 0.1 = 0.5 & \\
& gP(\hat{r}^{13}) = 0.4 + 0.4 - 0.2 = 0.6 & gP(\hat{r}^{14}) = 0.2 + 0.5 - 0.1 = 0.6 & 
\end{align*}

Therefore, the values of $gP$ and $gR$ are determined by their values on the set of join-irreducible elements.

\section{Rank-based Retrieval}
\label{sec:rank-based}
In this scenario, a judged run is an array of relevance degrees. Considering the \emph{swapping} operation individually will lead to a disconnected Hasse diagram. For example, the same result is always obtained when it is applied to $(\mathit{a}_1, \ldots, \mathit{a}_1)$; then, this element will not be comparable to any other judged run of $R(N)$. The same can be said for the \emph{projection} operation, which selects only specific positions in the ranking. On the other hand, the \emph{replacement} operation should be present in any combination since prioritizing relevant documents is a requirement of any retrieval system. 

Then, it is possible to consider the following combinations: (i) the \emph{projection} and \emph{replacement} operations jointly  and (ii) the \emph{replacement} operation individually. 

To combine the \emph{swapping} operation with other operations, it must be noted that the covering relation cannot contain the following pairs $\hat{r} \precdot \hat{s}$, $\hat{s} \precdot \hat{t}$ and $\hat{r} \precdot \hat{t}$ at the same time (since it is a transitive reduction; see Section \ref{sec:lattice}). In fact, when the \emph{swapping} operation between any pair of positions is combined with the \emph{replacement} operation in any position of the ranking, it leads to binary relations that do not satisfy this condition. For example, for $R(3)$ in the binary case, $\hat{r} = (\mathit{a}_1, \mathit{a}_0, \mathit{ a}_0)$ covers $\hat{0} = (\mathit{a}_0, \mathit{a}_0, \mathit{a}_0)$ when considering \emph{replacement} in the first position. However, $\hat{0} \precdot (\mathit{a}_0, \mathit{a}_1, \mathit{a}_0) \precdot \hat{r}$ when considering \emph{swapping} in the first two positions.

This means that to combine the \emph{swapping} and the \emph{replacement} operations, some restriction should be imposed on the \emph{replacement} property, for example, prioritizing the relevant documents only in some ranking positions. In this case, note that \emph{replacement} should be considered only in the last positions of the ranking since otherwise situations in which the covering relation is not a transitive reduction of its partial order could occur; the example in the previous paragraph illustrates this situation.

Thus, the combination of (iii) the \emph{swapping} with the \emph{replacement} operation only at the last position is also considered.

Finally, combining the three operations, i.e., \emph{swapping}, \emph{replacement} only at the last position and \emph{projection}, would lead to a poset whose covering relation is not a transitive reduction of its partial order. For example, for $R(3)$ in the binary case, $\hat {s} = (\mathit{a}_1, \mathit{a}_0, \mathit{a}_0)$ covers $\hat{r} = (\mathit{a}_0, \mathit{a}_1, \mathit{a}_0)$ when considering \emph{swapping} between the first and second positions. However, considering \emph{replacement} in the third position, $\hat{r} \precdot (\mathit{a}_0, \mathit{a}_1, \mathit{a}_1)= \hat{t}$, and when applying the \emph{projection} operation $\hat{t} \precdot \hat{s}$ would be obtained.

\subsection{Projection+Replacement [rank-based]}
Given a pair of judged runs, the \emph{projection} operation selects the more relevant part of the ranking, i.e., the highest-ranking position at which the two system runs differ. In addition, if the \emph{replacement} operation is also considered, the ranking with the highest \emph{cumulative mass of relevance} (or equivalently, the highest relevance degree\footnote{Considering the highest ranking position at which the two system runs differ, the difference between the \emph{cumulated mass of relevance} is determined by the relevance degree at that position.}) at that position is preferred. Formally:
\begin{quote}
\textbf{Projection+Replacement} [rank-based]

Let $\hat{r}$, $\hat{s} \in R(N)$ such that $\hat{r} \neq \hat{s}$; then, there exists $k = \min \{j \leq N : \hat{r}[j] \neq \hat{s}[j] \}$. Any pair of distinct system runs is ordered by: 
$$\hat{r} \preceq \hat{s} \Longleftrightarrow \hat{r}[k] \preccurlyeq \hat{s}[k]$$
\end{quote}
A system run is preferred to another if it has a higher relevance degree in the highest-rank position at which the two system runs differ. 

$(R(N), \preceq)$ is a totally ordered set or a chain\footnote{This partial order is the same as the ``Total Ordering'' (rank-based case) studied in \cite{ferrante2018general}, where it can be seen the linearity of the ordering.}. Thus, this is a trivial case as in the \textbf{projection+replacement} [set-based] total order.

\subsection{Replacement [rank-based]}
Considering the \emph{replacement} operation individually, the relevant documents must be prioritized, i.e., a system run is preferred to another if it has more \emph{cumulated mass of relevance}, for any relevance degree. 
Formally:
\begin{quote}
\textbf{Replacement} [rank-based]

Any pair of system runs, $\hat{r}, \hat{s} \in R(N)$, such that $\hat{r} \neq \hat{s}$, is ordered by: 
$$\hat{r} \preceq \hat{s} \Longleftrightarrow \big\vert\{i : \hat{r}[i] \succcurlyeq \mathit{a}_j\}\big\vert \leq \big\vert\{i : \hat{s}[i] \succcurlyeq \mathit{a}_j\}\big\vert, \forall j \in \{0, \ldots, c\}$$
\end{quote}
\begin{remark}
Considering the \emph{cumulative mass of relevance}, this ordering can be equivalently described with the following binary relation. Let $\hat{r}$, $\hat{s} \in R(N)$ be two judged runs:
$$\hat{r} \preceq \hat{s} \Longleftrightarrow \hat{r}[k] \preccurlyeq \hat{s}[k], \ \  \forall k \in \{1, \ldots, N\}$$
\end{remark}
A judged run is greater than another if, at each rank position, it has higher relevance degrees.

In this ordering not every judged run is comparable. For instance, for $N=3$ in the binary case, $\hat{r}=(\mathit{a}_1, \mathit{a}_0, \mathit{a}_0)$ and $\hat{s}=(\mathit{a}_0, \mathit{a}_1, \mathit{a}_0)$ are not comparable, since $\hat{s}[1] \preccurlyeq \hat{r}[1]$, but $\hat{r}[2] \preccurlyeq \hat{s}[2]$.

However, this ordering is a poset since $(REL, \preccurlyeq)$ is a totally ordered set. Given $\hat{r}, \hat{s}, \hat{t} \in R(N)$ and considering each position in the ranking, it is reflexive since $\hat{r}[k] \preccurlyeq \hat{r}[k]$; it is antisymmetric since if $\hat{r}[k] \preccurlyeq \hat{s}[k]$ and $\hat{s}[k] \preccurlyeq \hat{r}[k]$, then $\hat{r}[k] = \hat{s}[k]$; and it is transitive since if $\hat{r}[k] \preccurlyeq \hat{s}[k]$ and $\hat{s}[k] \preccurlyeq \hat{t}[k]$, then $\hat{r}[k] \preccurlyeq \hat{t}[k]$.

The join and meet operations have an explicit expression. 
\begin{proposition}
\label{prop:operations-rank}
Given $\hat{r} = (\hat{r}_1, \ldots, \hat{r}_N)$ and $\hat{s} = (\hat{s}_1, \ldots, \hat{s}_N)$ in $(R(N), \wedge, \vee, \preceq)$, with the \textbf{replacement} [rank-based] partial order, the meet and join can be expressed as:
\begin{align*}
\hat{r} \wedge \hat{s} & = \Big( \min \{ \hat{r}_1 , \hat{s}_1 \}, \ldots, \min \{ \hat{r}_N , \hat{s}_{N} \} \Big) \\
\hat{r} \vee \hat{s} & = \Big( \max \{ \hat{r}_1 , \hat{s}_1 \}, \ldots, \max \{ \hat{r}_N , \hat{s}_N \} \Big)
\end{align*}
\end{proposition}
In this case, $(R(N), \preceq)$ endowed with the meet and join operation is, trivially, a lattice since $R(N)$ is a finite set. It can be said more about the structure of $R(N)$ as it is shown in the following result.
\begin{proposition}
\label{prop:distrib-rank}
$(R(N), \wedge, \vee, \preceq)$, with the \textbf{replacement} [rank-based] partial order, is a distributive lattice.
\end{proposition}
In this context, a remarkable subset of judged runs is identified with the following result.
\begin{proposition}
\label{prop:irred-rank}
The join-irreducible elements of $(R(N), \wedge, \vee, \preceq)$, where $\preceq$ is the \textbf{replacement} [rank-based] ordering, are:
\begin{small}
\begin{equation*}
\begin{matrix}
(\mathit{a}_1, \mathit{a}_0, \ldots, \mathit{a}_0 ), & (\mathit{a}_0, \mathit{a}_1, \mathit{a}_0, \ldots, \mathit{a}_0 ), & \ldots, & (\mathit{a}_0, \mathit{a}_0, \ldots, \mathit{a}_1),  \\
(\mathit{a}_2, \mathit{a}_0, \ldots, \mathit{a}_0 ), & (\mathit{a}_0, \mathit{a}_2, \mathit{a}_0, \ldots, \mathit{a}_0 ), & \ldots, & (\mathit{a}_0, \mathit{a}_0, \ldots, \mathit{a}_2 ),  \\
\vdots & \vdots & \vdots & \vdots \\
(\mathit{a}_c, \mathit{a}_0, \ldots, \mathit{a}_0 ), & (\mathit{a}_0, \mathit{a}_c, \mathit{a}_0, \ldots, \mathit{a}_0), & \ldots, & (\mathit{a}_0, \mathit{a}_0, \ldots, \mathit{a}_c) 
\end{matrix}
\end{equation*}
\end{small}
\end{proposition}

\begin{figure}[bt]
  \centering
  %\scalebox{0.95}
  %{
  \includegraphics[width=\linewidth]{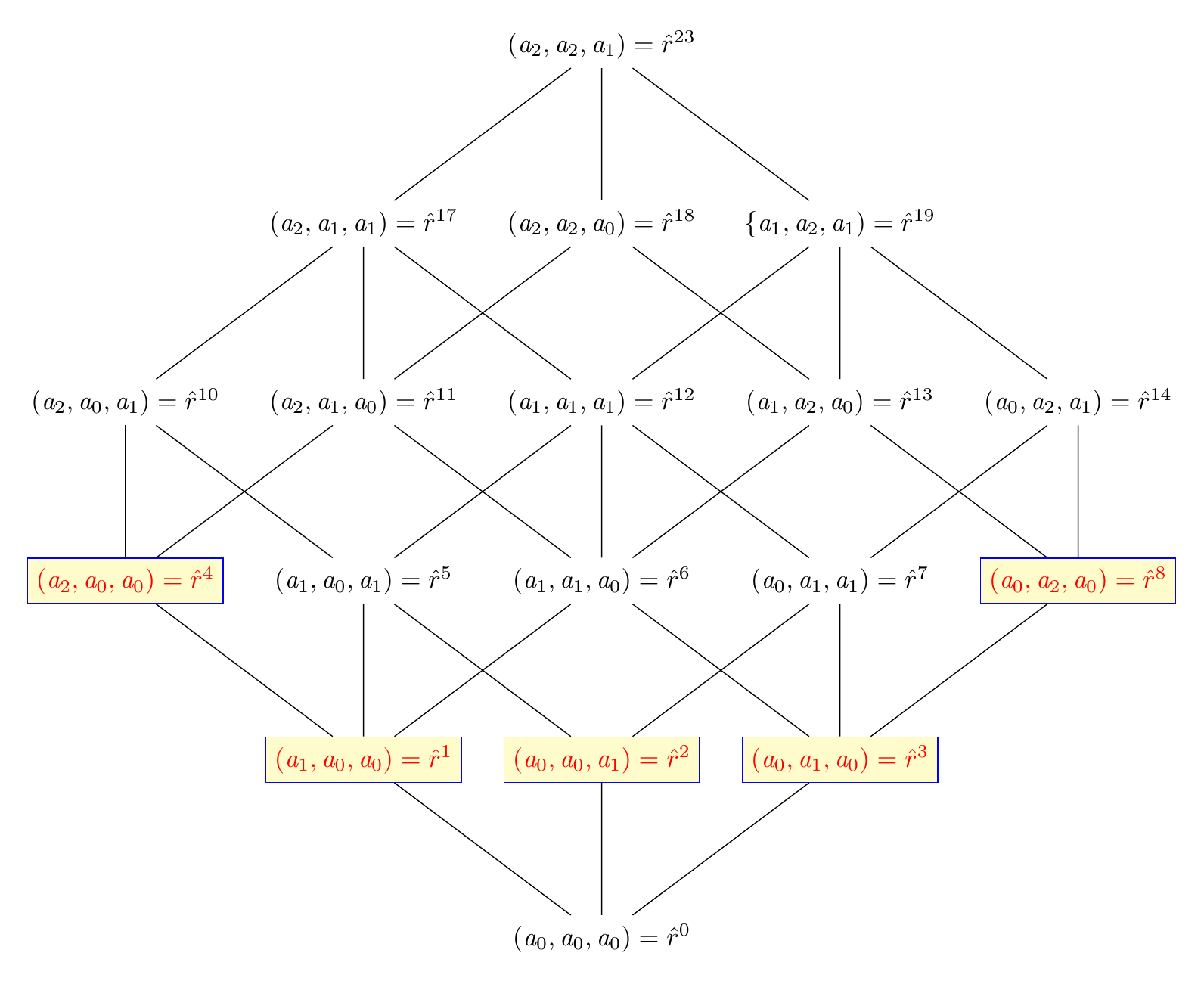}
  %}
  \caption{Hasse diagram for $c=2$ and $N=3$ in the rank-based retrieval, part 1}
	\label{fig:join-n3-i}
  \Description{Hasse diagram of the example ($c=2$ and $N=3$) in the rank-based retrieval, part 1.}
\end{figure}

\begin{figure}[bt]
  \centering
  %\scalebox{0.95}
  %{
  \includegraphics[width=\linewidth]{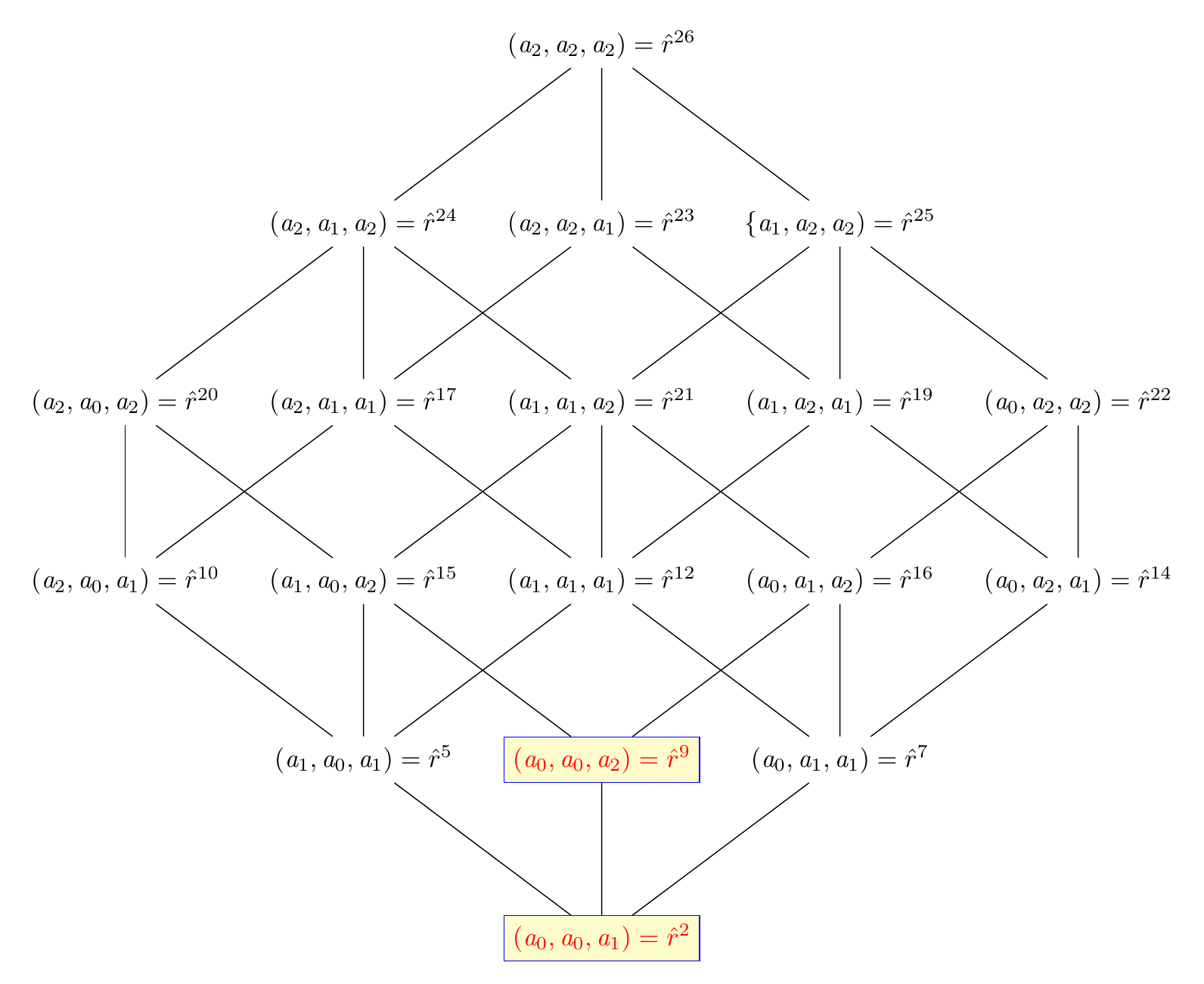}
  %}
  \caption{Hasse diagram for $c=2$ and $N=3$ in the rank-based retrieval, part 2}
	\label{fig:join-n3-ii}
  \Description{Hasse diagram of the example ($c=2$ and $N=3$) in the rank-based retrieval, part 2.}
\end{figure}

For example, for the case $N=3$ and three relevance degrees ($c=2$), the Hasse diagram has been split into Figures \ref{fig:join-n3-i} and \ref{fig:join-n3-ii}. It can be useful to visualize each figure in the three-dimensional space; thus, there are four cubes. Figure \ref{fig:join-n3-ii} should be placed on top of Figure \ref{fig:join-n3-i}, it can be seen that the top elements of Figure \ref{fig:join-n3-i} coincide with the bottom elements of Figure \ref{fig:join-n3-ii}. The join-irreducible elements have been marked in these figures.

The meet and join of two system runs can be graphically obtained in a similar manner as in the \textbf{replacement} [set-based] partial order. The structural properties of this partial order are obtained by applying Theorem \ref{th:expresion-join}; thus, every judged run can be expressed as a join of the join-irreducible elements.

\begin{figure}[bt]
  \centering
  \scalebox{1.05}
  {
  \includegraphics[width=\linewidth]{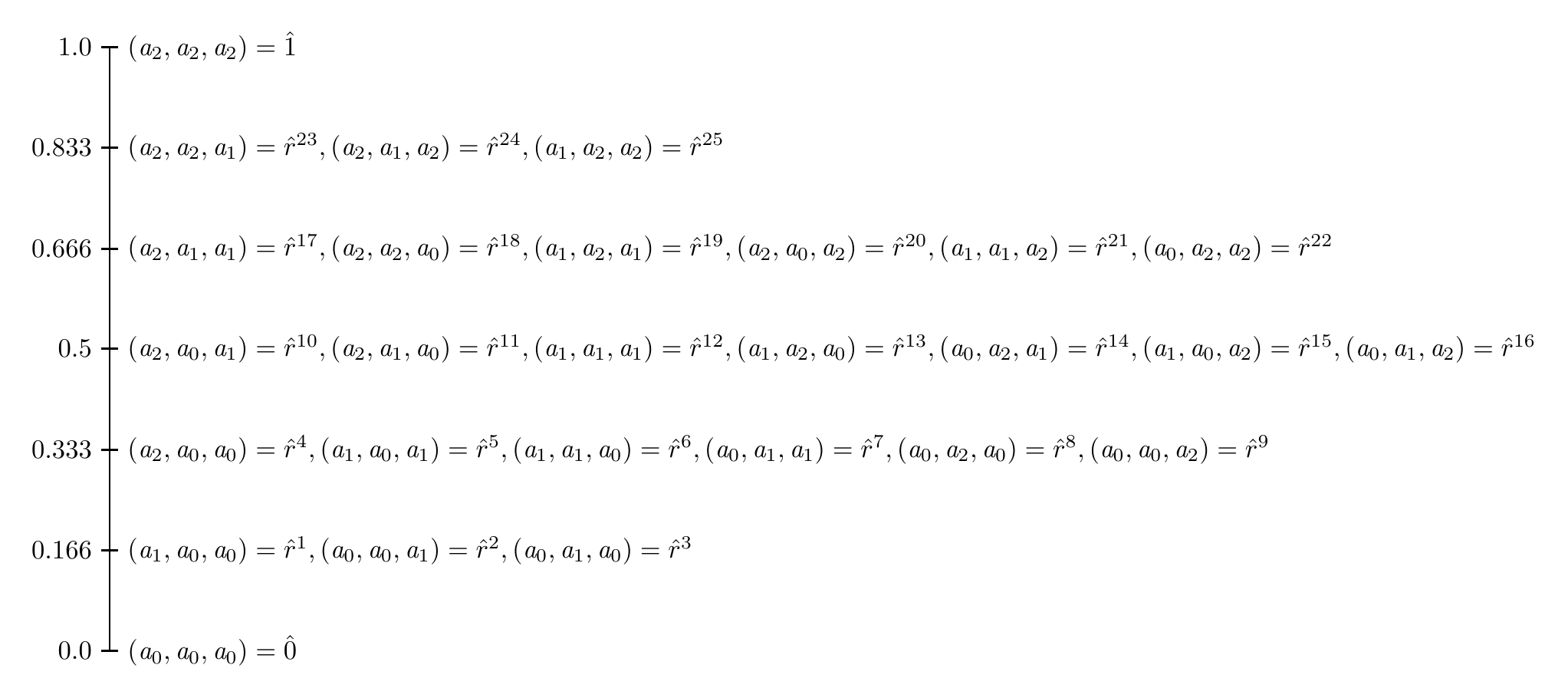}
  }
  \caption{Generalized Precision and Recall for $c=2$ and $N=3$ in the rank-based retrieval}
	\label{fig:prec-recall-rank}
  \Description{Values of the generalized Precision and Recall ($c=2$ and $N=3$) in the rank-based retrieval.}
\end{figure}

Once the properties of the empirical domain have been analysed, it is possible to study the properties of the evaluation metrics defined on it. Figure \ref{fig:prec-recall-rank} shows the numerical values reached by the generalized precision and recall (they are the same) for this example. These values have been obtained with the definition of $gP$ in Section \ref{sec:notation}.

In this scenario, a noteworthy property of the generalized precision and recall, $gRBP$ and $DCG_b$ is the following result.
\begin{proposition}
\label{prop:valuation-rank}
The generalized precision and recall, $gRBP$ and $DCG_b$ are valuations, when they are defined on $(R(N), \wedge, \vee, \preceq)$, where $\preceq$ is the \textbf{replacement} [rank-based] partial order.
\end{proposition}

The main result of this subsection is obtained from Propositions \ref{prop:distrib-rank} and \ref{prop:valuation-rank}, which allow to apply Theorem \ref{th:valuation-distributive} to this partial order. Thus, the values reached by $gP$, $gR$, $gRBP$ and $DCG_b$ are determined by the values reached on the set of join-irreducible elements.

Let us now look at some practical aspects of calculating these values. By means of Proposition \ref{prop:valuation-set}, the numerical values of these evaluation metrics can be computed with the characterization property of valuations: $v(\hat{r} \vee \hat{s}) = v(\hat{r}) + v(\hat{s}) - v(\hat{r} \wedge \hat{s})$. Note that the last operand of the second part of the equality contains a meet of two join-irreducible elements. Considering Propositions \ref{prop:operations-rank} and \ref{prop:irred-rank}, it can easily be seen that the meet of any pair of join-irreducible elements is join-irreducible. Thus, the characterization of valuations is sufficient to determine any numerical value.

For example, in Figure \ref{fig:prec-recall-rank}, it is shown that $gP(\hat{r}^5) = 0.333$. This value can also be obtained from the values of $gP$ in the set of join-irreducible elements. Making use of Theorem \ref{th:expresion-join} and Proposition \ref{prop:valuation-rank}, $gP(\hat{r}^5) = gP(\hat{r}^1 \vee \hat{r}^2)  = gP(\hat{r}^1) + gP(\hat{r}^2) - gP(\hat{r}^1 \wedge \hat{r}^2) = 0.166 + 0.166 - gP(\hat{0}) = 0.333 - 0.0 = 0.333$.

Another example where a system run is expressed with three join-irreducible elements is the following
$$gP(\hat{r}^{17}) = gP(\hat{r}_4 \vee \hat{r}^2 \vee \hat{r}^3) = gP(\hat{r}^4) +gP(\hat{r}^2 \vee \hat{r}^3) - gP(\hat{r}^4 \wedge (\hat{r}^2 \vee \hat{r}^3)) .$$

As the Hasse diagram for the \textbf{replacement} [rank-based] ordering is a distributive lattice, the argument of the last operand, in the previous expression is:
$$\hat{r}^4 \wedge (\hat{r}^2 \vee \hat{r}^3) = (\hat{r}^4 \wedge \hat{r}^2) \vee (\hat{r}^4 \wedge \hat{r}^3)= \hat{0} \vee \hat{0} = \hat{0} .$$

Considering that $gP(\hat{0}) = 0$, Theorem \ref{th:expresion-join} and Proposition \ref{prop:valuation-rank}:
\begin{equation*}
\begin{matrix}
gP(\hat{r}^{17}) = 0.333 +gP(\hat{r}^2 \vee \hat{r}^3) - 0= 0.333 + gP(\hat{r}^2) + gP(\hat{r}^3) - \\
- gP(\hat{r}^2 \wedge \hat{r}^3) = 0.333 + 0.166 + 0.166-0 = 0.666 . \hfill
\end{matrix}
\end{equation*}

The $gRBP$ and $DCG_b$ evaluation metrics benefit from the same properties, we omit the checks for space limitations.

\subsection{Replacement+Swapping [rank-based]}
Considering the \emph{replacement} operation, the relevant documents must be prioritized, i.e., a system run is preferred to another, if it has more \emph{cumulated mass of relevance} for any relevance degree. In addition, when the \emph{swapping} operation is considered, this prioritization is required at any position of the ranking since it is possible to interchange pairs of documents.

Thus, a system run is preferred to another if, for any relevance degree, it has more \emph{cumulated mass of relevance} at any position of the ranking than the other. 
Formally:
\begin{quote}
\textbf{Replacement+Swapping} [rank-based]

Given two system runs $\hat{r}$, $\hat{s} \in R(N)$,
\begin{align*}
& \hat{r} \preceq \hat{s} \Longleftrightarrow \big\vert\{i \leq k : \hat{r}[i] \succcurlyeq \mathit{a}_j\}\big\vert \leq \big\vert\{i \leq k : \hat{s}[i] \succcurlyeq \mathit{a}_j\}\big\vert \\
& \forall j \in \{0, \ldots, c\}, \forall k \in \{1, \ldots, N\}
\end{align*}
\end{quote}
A run is considered larger than another one when, at each rank position, it has more relevant documents than the other up to that rank for every relevance degree.

$(R(N), \preceq)$ is a partial order\footnote{This partial order is the same as the ``Partial Ordering'' (rank-based case) studied in \cite{ferrante2018general}, where it can be seen this result.}, where not every pair of runs is comparable. For example, for $c=2$, $N = 3$, $(\mathit{a}_2, \mathit{a}_2, \mathit{a}_1)$, and $(\mathit{a}_2, \mathit{a}_2, \mathit{a}_0)$ are not comparable.

Moreover, considering $\hat{r} = (\mathit{a}_2, \mathit{a}_1, \mathit{a}_1)$, and $\hat{s} = (\mathit{a}_2, \mathit{a}_2, \mathit{a}_1)$, $\hat{r} \preceq \hat{s}$ and the interval $[\hat{r}, \hat{s}]$ is a sublattice of $R(N)$, whose Hasse diagram is an $N_5$ \cite{ferrante2018general}. Thus, by Theorem \ref{th:distributive_n_m}, $(R(N), \wedge, \vee, \preceq)$ is not a distributive lattice. Consequently, not every element can be expressed in a unique manner as a join of join-irreducible elements. Thus, the values of the evaluation metrics, defined on this partial order, do not necessarily are determined by the values on the join-irreducible elements. 

\section{Conclusions and Applicability}
\label{sec:conclusion}
Building on some orderings on the set of ranked lists of documents, it has been verified that the properties of some IR evaluation metrics, namely, precision, recall, $DCG$ and $RBP$ do not depend only on its analytical expression. The empirical domain where they are defined play an important role. When the relevant documents are prioritised, these evaluation metrics are completely determined from their values on the subset of join-irreducible elements. However, this is not the case, when the swapping of documents is considered. 

Our contribution is three fold: (i) the proposal grounds the behaviour of IR evaluation metrics in a representative and well understood basis of three retrieval axioms; (ii) the structural properties of the orderings are determined, in the context of lattice theory, identifying the join-irreducible elements, which have some resemblance to the concept of basis in vector space; and (iii) some numerical properties of retrieval metrics have been explained in terms of IR axioms. They can be generalized to some Carterette's models \cite{carterette2011system} and some Moffat's static user models \cite{moffat2013users},  in which the conditional continuation probabilities are a function of the rank position alone.

The results of this paper have some applicability to the study of the behaviour of retrieval metrics; for instance, it has been shown that, in the \textbf{replacement} [rank-based] ordering, the total number of possible system runs of length $N$ with $c$ relevance degrees is $c^N$ and the number of join-irreducible elements is $c \cdot N$. At practical level, in a rank-based retrieval of $100$ documents and $3$ relevance degrees, the values of precision, recall, $RBP$ or $DCG$ on any of the $3^{100}$ possible system output runs are determined by only $300$ values (see Theorem \ref{th:valuation-distributive}). This result is also applicable to any other IR evaluation metric verifying the valuation condition; thus, it allows to define evaluation metrics without knowing their analytical expression. It is sufficient to assign $300$ values to the join-irreducible elements. In this way, it is possible to define fine-tuned IR evaluation metrics that prioritise one type of system runs over others, by assigning a higher value to the join-irreducible elements involved in those system runs.

Another practical application is the understanding of the metric performance and the determination of bounds. Suppose that, in the rank-based retrieval of Figures \ref{fig:join-n3-i} and \ref{fig:join-n3-ii}, a system run can retrieve one document of relevance $\mathit{a}_2$ in the first position (i.e., $\hat{r}^4$) or $2$ documents of relevance $\mathit{a}_1$ in the first and second positions (i.e., $\hat{r}^6$). Which of these two system runs should have a higher value for an IR evaluation metric? It seems that the system run with one document of relevance $\mathit{a}_2$ in the second position (i.e., $\hat{r}^8$) should have a lower value than $\hat{r}^4$; and according to Figure \ref{fig:join-n3-i}, $\hat{r}^6$ should have some value in between. $RBP$ verifies this property, showing its granularity; however, precision, recall and $DCG$ assign the same value to the three judged runs. On the other hand, if the best achievement of a system is to retrieve one document of relevance $\mathit{a}_2$ or two documents of relevance $\mathit{a}_1$, then the value of an IR evaluation metric on the join of $\hat{r}^4$ and $\hat{r}^6$ is a bound of system performance. 

At theoretical level, determining the structural properties of the total\slash partial orderings derived from ranking axioms have been one of the key points of this paper. It enables to apply many results of lattice theory to the set of rankings. Considering the close relation between lattices and metric spaces \cite{birkhoff1940lattice}, it is possible to classify the IR evaluation measures into metric and\slash or pseudo-metric mappings. In addition, taking advantage of the concept of isotone mapping and interval, it can be determined the possible ordinal and interval scales defined on these partial orders. It may open up new ways to understand ongoing debates about questions like ``can we use MRR?'' \cite{fuhr2021proof,sakai2021fuhr, ferrante2021towards}. 

This paper has described a procedure to convert ranking axioms or operations into total\slash partial orders. They can be understood as ideal scenarios where some desirable properties are considered. However, these orderings cannot be considered the only ones that can be defined in the set of rankings. Considering different ranking axioms or operations could lead to other ideal scenarios which will be interesting objects to be analysed. 

%This paper has described a procedure to convert ranking axioms or operations into total\slash partial orders. They can be understood as ideal scenarios where some desirable properties are considered. Although, there is at least a total order and a partial order in each type of retrieval (set-based\slash rank-based); these orderings cannot be considered as the only ones that can be defined in the set of rankings. Thus, categorical statements cannot be made on the basis of these orderings alone, for instance, the sentence ``that evaluation metric posses certain property'' should be specified with the scenario in which it is applied and the assumptions on which it is based. Considering different ranking axioms or operations could lead to other ideal scenarios which will be interesting objects to be analysed. 

\begin{acks}
The author thanks the anonymous reviewers for their valuable comments and suggestions.
\end{acks}

\bibliographystyle{ACM-Reference-Format}
\bibliography{thebibliography}


\begin{thebibliography}{55}

%%% the Web address.

\ifx \showCODEN    \undefined \def \showCODEN     #1{\unskip}     \fi
\ifx \showDOI      \undefined \def \showDOI       #1{#1}\fi
\ifx \showISBNx    \undefined \def \showISBNx     #1{\unskip}     \fi
\ifx \showISBNxiii \undefined \def \showISBNxiii  #1{\unskip}     \fi
\ifx \showISSN     \undefined \def \showISSN      #1{\unskip}     \fi
\ifx \showLCCN     \undefined \def \showLCCN      #1{\unskip}     \fi
\ifx \shownote     \undefined \def \shownote      #1{#1}          \fi
\ifx \showarticletitle \undefined \def \showarticletitle #1{#1}   \fi
\ifx \showURL      \undefined \def \showURL       {\relax}        \fi
% invisible to TeX
\providecommand\bibfield[2]{#2}
\providecommand\bibinfo[2]{#2}
\providecommand\natexlab[1]{#1}
\providecommand\showeprint[2][]{arXiv:#2}

\bibitem[Amig{\'o} et~al\mbox{.}(2017a)]%
        {amigo2017axiomatic}
\bibfield{author}{\bibinfo{person}{Enrique Amig{\'o}}, \bibinfo{person}{Hui
  Fang}, \bibinfo{person}{Stefano Mizzaro}, {and} \bibinfo{person}{ChengXiang
  Zhai}.} \bibinfo{year}{2017}\natexlab{a}.
\newblock \showarticletitle{Axiomatic thinking for information retrieval: And
  related tasks}. In \bibinfo{booktitle}{\emph{Proceedings of the 40th
  international ACM SIGIR conference on research and development in information
  retrieval}}. \bibinfo{pages}{1419--1420}.
\newblock


\bibitem[Amig{\'o} et~al\mbox{.}(2020)]%
        {amigo2020axiomatic}
\bibfield{author}{\bibinfo{person}{Enrique Amig{\'o}}, \bibinfo{person}{Hui
  Fang}, \bibinfo{person}{Stefano Mizzaro}, {and} \bibinfo{person}{Chengxiang
  Zhai}.} \bibinfo{year}{2020}\natexlab{}.
\newblock \showarticletitle{Axiomatic thinking for information retrieval:
  introduction to special issue}.
\newblock \bibinfo{journal}{\emph{Information Retrieval Journal}}
  \bibinfo{volume}{23}, \bibinfo{number}{3} (\bibinfo{year}{2020}),
  \bibinfo{pages}{187--190}.
\newblock


\bibitem[Amig{\'o} et~al\mbox{.}(2017b)]%
        {amigo2017axiomatic2}
\bibfield{author}{\bibinfo{person}{Enrique Amig{\'o}},
  \bibinfo{person}{Fernando Giner}, \bibinfo{person}{Julio Gonzalo}, {and}
  \bibinfo{person}{Felisa Verdejo}.} \bibinfo{year}{2017}\natexlab{b}.
\newblock \showarticletitle{An axiomatic account of similarity}. In
  \bibinfo{booktitle}{\emph{Proceedings of the SIGIR’17 Workshop on Axiomatic
  Thinking for Information Retrieval and Related Tasks (ATIR)}}.
\newblock


\bibitem[Amig{\'o} et~al\mbox{.}(2009)]%
        {amigo2009comparison}
\bibfield{author}{\bibinfo{person}{Enrique Amig{\'o}}, \bibinfo{person}{Julio
  Gonzalo}, \bibinfo{person}{Javier Artiles}, {and} \bibinfo{person}{Felisa
  Verdejo}.} \bibinfo{year}{2009}\natexlab{}.
\newblock \showarticletitle{A comparison of extrinsic clustering evaluation
  metrics based on formal constraints}.
\newblock \bibinfo{journal}{\emph{Information retrieval}} \bibinfo{volume}{12},
  \bibinfo{number}{4} (\bibinfo{year}{2009}), \bibinfo{pages}{461--486}.
\newblock


\bibitem[Amig{\'o} et~al\mbox{.}(2013)]%
        {amigo2013general}
\bibfield{author}{\bibinfo{person}{Enrique Amig{\'o}}, \bibinfo{person}{Julio
  Gonzalo}, {and} \bibinfo{person}{Felisa Verdejo}.}
  \bibinfo{year}{2013}\natexlab{}.
\newblock \showarticletitle{A general evaluation measure for document
  organization tasks}. In \bibinfo{booktitle}{\emph{Proceedings of the 36th
  international ACM SIGIR conference on Research and development in information
  retrieval}}. \bibinfo{pages}{643--652}.
\newblock


\bibitem[Amig{\'o} and Mizzaro(2020)]%
        {amigo2020nature}
\bibfield{author}{\bibinfo{person}{Enrique Amig{\'o}} {and}
  \bibinfo{person}{Stefano Mizzaro}.} \bibinfo{year}{2020}\natexlab{}.
\newblock \showarticletitle{On the nature of information access evaluation
  metrics: a unifying framework}.
\newblock \bibinfo{journal}{\emph{Information Retrieval Journal}}
  \bibinfo{volume}{23}, \bibinfo{number}{3} (\bibinfo{year}{2020}),
  \bibinfo{pages}{318--386}.
\newblock


\bibitem[Amig{\'o} et~al\mbox{.}(2018)]%
        {amigo2018axiomatic}
\bibfield{author}{\bibinfo{person}{Enrique Amig{\'o}}, \bibinfo{person}{Damiano
  Spina}, {and} \bibinfo{person}{Jorge Carrillo-de Albornoz}.}
  \bibinfo{year}{2018}\natexlab{}.
\newblock \showarticletitle{An axiomatic analysis of diversity evaluation
  metrics: Introducing the rank-biased utility metric}. In
  \bibinfo{booktitle}{\emph{The 41st International ACM SIGIR Conference on
  Research \& Development in Information Retrieval}}.
  \bibinfo{pages}{625--634}.
\newblock


\bibitem[Birkhoff(1940)]%
        {birkhoff1940lattice}
\bibfield{author}{\bibinfo{person}{Garrett Birkhoff}.}
  \bibinfo{year}{1940}\natexlab{}.
\newblock \bibinfo{booktitle}{\emph{Lattice theory}}.
  Vol.~\bibinfo{volume}{25}.
\newblock \bibinfo{publisher}{American Mathematical Soc.}
\newblock


\bibitem[Blair(1979)]%
        {van1979information}
\bibfield{author}{\bibinfo{person}{David~C. Blair}.}
  \bibinfo{year}{1979}\natexlab{}.
\newblock \showarticletitle{Information Retrieval, 2nd ed. {C.J.} Van
  Rijsbergen. London: Butterworths; 1979: 208 pp. Price: {\textdollar}32.50}.
\newblock \bibinfo{journal}{\emph{{JASIS}}} \bibinfo{volume}{30},
  \bibinfo{number}{6} (\bibinfo{year}{1979}), \bibinfo{pages}{374--375}.
\newblock
\urldef\tempurl%
\url{https://doi.org/10.1002/asi.4630300621}
\showDOI{\tempurl}


\bibitem[Blyth(2005)]%
        {blyth2005distributive}
\bibfield{author}{\bibinfo{person}{TS Blyth}.} \bibinfo{year}{2005}\natexlab{}.
\newblock \showarticletitle{Distributive lattices}.
\newblock \bibinfo{journal}{\emph{Lattices and Ordered Algebraic Structures}}
  (\bibinfo{year}{2005}), \bibinfo{pages}{65--76}.
\newblock


\bibitem[Bollmann(1984)]%
        {bollmann1984two}
\bibfield{author}{\bibinfo{person}{Peter Bollmann}.}
  \bibinfo{year}{1984}\natexlab{}.
\newblock \showarticletitle{Two axioms for evaluation measures in information
  retrieval}. In \bibinfo{booktitle}{\emph{SIGIR}}, Vol.~\bibinfo{volume}{84}.
  Citeseer, \bibinfo{pages}{233--245}.
\newblock


\bibitem[Bollmann and Cherniavsky(1980)]%
        {bollmann1980measurement}
\bibfield{author}{\bibinfo{person}{Peter Bollmann} {and}
  \bibinfo{person}{Vladimir~S. Cherniavsky}.} \bibinfo{year}{1980}\natexlab{}.
\newblock \showarticletitle{Measurement-theoretical investigation of the
  MZ-metric}. In \bibinfo{booktitle}{\emph{Proceedings of the 3rd annual ACM
  conference on Research and development in information retrieval}}. Citeseer,
  \bibinfo{pages}{256--267}.
\newblock


\bibitem[Busin and Mizzaro(2013)]%
        {busin2013axiometrics}
\bibfield{author}{\bibinfo{person}{Luca Busin} {and} \bibinfo{person}{Stefano
  Mizzaro}.} \bibinfo{year}{2013}\natexlab{}.
\newblock \showarticletitle{Axiometrics: An axiomatic approach to information
  retrieval effectiveness metrics}. In \bibinfo{booktitle}{\emph{Proceedings of
  the 2013 Conference on the Theory of Information Retrieval}}.
  \bibinfo{pages}{22--29}.
\newblock


\bibitem[Carterette(2011)]%
        {carterette2011system}
\bibfield{author}{\bibinfo{person}{Ben Carterette}.}
  \bibinfo{year}{2011}\natexlab{}.
\newblock \showarticletitle{System effectiveness, user models, and user
  utility: a conceptual framework for investigation}. In
  \bibinfo{booktitle}{\emph{Proceedings of the 34th international ACM SIGIR
  conference on Research and development in information retrieval}}.
  \bibinfo{pages}{903--912}.
\newblock


\bibitem[Cleverdon(1967)]%
        {cleverdon1967cranfield}
\bibfield{author}{\bibinfo{person}{Cyril Cleverdon}.}
  \bibinfo{year}{1967}\natexlab{}.
\newblock \showarticletitle{The Cranfield tests on index language devices}. In
  \bibinfo{booktitle}{\emph{Aslib proceedings}}, Vol.~\bibinfo{volume}{19}. MCB
  UP Ltd, \bibinfo{pages}{173--194}.
\newblock


\bibitem[Clinchant and Gaussier(2011)]%
        {clinchant2011document}
\bibfield{author}{\bibinfo{person}{St{\'e}phane Clinchant} {and}
  \bibinfo{person}{Eric Gaussier}.} \bibinfo{year}{2011}\natexlab{}.
\newblock \showarticletitle{Is document frequency important for PRF?}. In
  \bibinfo{booktitle}{\emph{Conference on the theory of information
  retrieval}}. Springer, \bibinfo{pages}{89--100}.
\newblock


\bibitem[Clinchant and Gaussier(2013)]%
        {clinchant2013theoretical}
\bibfield{author}{\bibinfo{person}{St{\'e}phane Clinchant} {and}
  \bibinfo{person}{Eric Gaussier}.} \bibinfo{year}{2013}\natexlab{}.
\newblock \showarticletitle{A theoretical analysis of pseudo-relevance feedback
  models}. In \bibinfo{booktitle}{\emph{Proceedings of the 2013 Conference on
  the Theory of Information Retrieval}}. \bibinfo{pages}{6--13}.
\newblock


\bibitem[Dom(2012)]%
        {dom2012information}
\bibfield{author}{\bibinfo{person}{Byron~E Dom}.}
  \bibinfo{year}{2012}\natexlab{}.
\newblock \showarticletitle{An information-theoretic external cluster-validity
  measure}.
\newblock \bibinfo{journal}{\emph{arXiv preprint arXiv:1301.0565}}
  (\bibinfo{year}{2012}).
\newblock


\bibitem[Fang(2007)]%
        {fang2007axiomatic}
\bibfield{author}{\bibinfo{person}{Hui Fang}.} \bibinfo{year}{2007}\natexlab{}.
\newblock \bibinfo{booktitle}{\emph{An axiomatic approach to information
  retrieval}}.
\newblock \bibinfo{type}{{T}echnical {R}eport}.
\newblock


\bibitem[Fang et~al\mbox{.}(2004)]%
        {fang2004formal}
\bibfield{author}{\bibinfo{person}{Hui Fang}, \bibinfo{person}{Tao Tao}, {and}
  \bibinfo{person}{ChengXiang Zhai}.} \bibinfo{year}{2004}\natexlab{}.
\newblock \showarticletitle{A formal study of information retrieval
  heuristics}. In \bibinfo{booktitle}{\emph{Proceedings of the 27th annual
  international ACM SIGIR conference on Research and development in information
  retrieval}}. \bibinfo{pages}{49--56}.
\newblock


\bibitem[Fang et~al\mbox{.}(2011)]%
        {fang2011diagnostic}
\bibfield{author}{\bibinfo{person}{Hui Fang}, \bibinfo{person}{Tao Tao}, {and}
  \bibinfo{person}{Chengxiang Zhai}.} \bibinfo{year}{2011}\natexlab{}.
\newblock \showarticletitle{Diagnostic evaluation of information retrieval
  models}.
\newblock \bibinfo{journal}{\emph{ACM Transactions on Information Systems
  (TOIS)}} \bibinfo{volume}{29}, \bibinfo{number}{2} (\bibinfo{year}{2011}),
  \bibinfo{pages}{1--42}.
\newblock


\bibitem[Fang and Zhai(2005)]%
        {fang2005exploration}
\bibfield{author}{\bibinfo{person}{Hui Fang} {and} \bibinfo{person}{ChengXiang
  Zhai}.} \bibinfo{year}{2005}\natexlab{}.
\newblock \showarticletitle{An exploration of axiomatic approaches to
  information retrieval}. In \bibinfo{booktitle}{\emph{Proceedings of the 28th
  annual international ACM SIGIR conference on Research and development in
  information retrieval}}. \bibinfo{pages}{480--487}.
\newblock


\bibitem[Fang and Zhai(2006)]%
        {fang2006semantic}
\bibfield{author}{\bibinfo{person}{Hui Fang} {and} \bibinfo{person}{ChengXiang
  Zhai}.} \bibinfo{year}{2006}\natexlab{}.
\newblock \showarticletitle{Semantic term matching in axiomatic approaches to
  information retrieval}. In \bibinfo{booktitle}{\emph{Proceedings of the 29th
  annual international ACM SIGIR conference on Research and development in
  information retrieval}}. \bibinfo{pages}{115--122}.
\newblock


\bibitem[Ferrante et~al\mbox{.}(2021)]%
        {ferrante2021towards}
\bibfield{author}{\bibinfo{person}{Marco Ferrante}, \bibinfo{person}{Nicola
  Ferro}, {and} \bibinfo{person}{Norbert Fuhr}.}
  \bibinfo{year}{2021}\natexlab{}.
\newblock \showarticletitle{Towards Meaningful Statements in IR Evaluation.
  Mapping Evaluation Measures to Interval Scales}.
\newblock \bibinfo{journal}{\emph{arXiv preprint arXiv:2101.02668}}
  (\bibinfo{year}{2021}).
\newblock


\bibitem[Ferrante et~al\mbox{.}(2015)]%
        {ferrante2015towards}
\bibfield{author}{\bibinfo{person}{Marco Ferrante}, \bibinfo{person}{Nicola
  Ferro}, {and} \bibinfo{person}{Maria Maistro}.}
  \bibinfo{year}{2015}\natexlab{}.
\newblock \showarticletitle{Towards a formal framework for utility-oriented
  measurements of retrieval effectiveness}. In
  \bibinfo{booktitle}{\emph{Proceedings of the 2015 International Conference on
  The Theory of Information Retrieval}}. \bibinfo{pages}{21--30}.
\newblock


\bibitem[Ferrante et~al\mbox{.}(2018)]%
        {ferrante2018general}
\bibfield{author}{\bibinfo{person}{Marco Ferrante}, \bibinfo{person}{Nicola
  Ferro}, {and} \bibinfo{person}{Silvia Pontarollo}.}
  \bibinfo{year}{2018}\natexlab{}.
\newblock \showarticletitle{A general theory of IR evaluation measures}.
\newblock \bibinfo{journal}{\emph{IEEE Transactions on Knowledge and Data
  Engineering}} \bibinfo{volume}{31}, \bibinfo{number}{3}
  (\bibinfo{year}{2018}), \bibinfo{pages}{409--422}.
\newblock


\bibitem[Fuhr(2021)]%
        {fuhr2021proof}
\bibfield{author}{\bibinfo{person}{Norbert Fuhr}.}
  \bibinfo{year}{2021}\natexlab{}.
\newblock \showarticletitle{Proof by experimentation? towards better IR
  research}. In \bibinfo{booktitle}{\emph{ACM SIGIR Forum}},
  Vol.~\bibinfo{volume}{54}. ACM New York, NY, USA, \bibinfo{pages}{1--4}.
\newblock


\bibitem[Gr{\"a}tzer(2002)]%
        {gratzer2002general}
\bibfield{author}{\bibinfo{person}{George Gr{\"a}tzer}.}
  \bibinfo{year}{2002}\natexlab{}.
\newblock \bibinfo{booktitle}{\emph{General lattice theory}}.
\newblock \bibinfo{publisher}{Springer Science \& Business Media}.
\newblock


\bibitem[Hagen et~al\mbox{.}(2016)]%
        {hagen2016axiomatic}
\bibfield{author}{\bibinfo{person}{Matthias Hagen}, \bibinfo{person}{Michael
  V{\"o}lske}, \bibinfo{person}{Steve G{\"o}ring}, {and} \bibinfo{person}{Benno
  Stein}.} \bibinfo{year}{2016}\natexlab{}.
\newblock \showarticletitle{Axiomatic result re-ranking}. In
  \bibinfo{booktitle}{\emph{Proceedings of the 25th ACM International on
  Conference on Information and Knowledge Management}}.
  \bibinfo{pages}{721--730}.
\newblock


\bibitem[Huibers(1996)]%
        {huibers1996axiomatic}
\bibfield{author}{\bibinfo{person}{Theodorus Wilhelmus~Charles Huibers}.}
  \bibinfo{year}{1996}\natexlab{}.
\newblock \emph{\bibinfo{title}{An axiomatic theory for information
  retrieval}}.
\newblock \bibinfo{thesistype}{Ph.\,D. Dissertation}.
\newblock


\bibitem[Karimzadehgan and Zhai(2012)]%
        {karimzadehgan2012axiomatic}
\bibfield{author}{\bibinfo{person}{Maryam Karimzadehgan} {and}
  \bibinfo{person}{ChengXiang Zhai}.} \bibinfo{year}{2012}\natexlab{}.
\newblock \showarticletitle{Axiomatic analysis of translation language model
  for information retrieval}. In \bibinfo{booktitle}{\emph{European Conference
  on Information Retrieval}}. Springer, \bibinfo{pages}{268--280}.
\newblock


\bibitem[Kek{\"a}l{\"a}inen and J{\"a}rvelin(2002)]%
        {kekalainen2002using}
\bibfield{author}{\bibinfo{person}{Jaana Kek{\"a}l{\"a}inen} {and}
  \bibinfo{person}{Kalervo J{\"a}rvelin}.} \bibinfo{year}{2002}\natexlab{}.
\newblock \showarticletitle{Using graded relevance assessments in IR
  evaluation}.
\newblock \bibinfo{journal}{\emph{Journal of the American Society for
  Information Science and Technology}} \bibinfo{volume}{53},
  \bibinfo{number}{13} (\bibinfo{year}{2002}), \bibinfo{pages}{1120--1129}.
\newblock


\bibitem[Krantz et~al\mbox{.}(1971)]%
        {krantz1971foundations}
\bibfield{author}{\bibinfo{person}{David Krantz}, \bibinfo{person}{Duncan
  Luce}, \bibinfo{person}{Patrick Suppes}, {and} \bibinfo{person}{Amos
  Tversky}.} \bibinfo{year}{1971}\natexlab{}.
\newblock \showarticletitle{Foundations of measurement, Vol. I: Additive and
  polynomial representations}.
\newblock  (\bibinfo{year}{1971}).
\newblock


\bibitem[Krantz(1989)]%
        {krantz1989foundations}
\bibfield{author}{\bibinfo{person}{David~H Krantz}.}
  \bibinfo{year}{1989}\natexlab{}.
\newblock \showarticletitle{Foundations of Measurement. Vol. II. Geometrical,
  Threshold and Probabilistic Representations}.
\newblock  (\bibinfo{year}{1989}).
\newblock


\bibitem[Maddalena and Mizzaro(2014)]%
        {maddalena2014axiometrics}
\bibfield{author}{\bibinfo{person}{Eddy Maddalena} {and}
  \bibinfo{person}{Stefano Mizzaro}.} \bibinfo{year}{2014}\natexlab{}.
\newblock \showarticletitle{Axiometrics: Axioms of Information Retrieval
  Effectiveness Metrics.}. In \bibinfo{booktitle}{\emph{EVIA@ NTCIR}}.
\newblock


\bibitem[Meil{\u{a}}(2007)]%
        {meilua2007comparing}
\bibfield{author}{\bibinfo{person}{Marina Meil{\u{a}}}.}
  \bibinfo{year}{2007}\natexlab{}.
\newblock \showarticletitle{Comparing clusterings—an information based
  distance}.
\newblock \bibinfo{journal}{\emph{Journal of multivariate analysis}}
  \bibinfo{volume}{98}, \bibinfo{number}{5} (\bibinfo{year}{2007}),
  \bibinfo{pages}{873--895}.
\newblock


\bibitem[Moffat(2013)]%
        {moffat2013seven}
\bibfield{author}{\bibinfo{person}{Alistair Moffat}.}
  \bibinfo{year}{2013}\natexlab{}.
\newblock \showarticletitle{Seven numeric properties of effectiveness metrics}.
  In \bibinfo{booktitle}{\emph{Asia Information Retrieval Symposium}}.
  Springer, \bibinfo{pages}{1--12}.
\newblock


\bibitem[Moffat et~al\mbox{.}(2013)]%
        {moffat2013users}
\bibfield{author}{\bibinfo{person}{Alistair Moffat}, \bibinfo{person}{Paul
  Thomas}, {and} \bibinfo{person}{Falk Scholer}.}
  \bibinfo{year}{2013}\natexlab{}.
\newblock \showarticletitle{Users versus models: What observation tells us
  about effectiveness metrics}. In \bibinfo{booktitle}{\emph{Proceedings of the
  22nd ACM international conference on Information \& Knowledge Management}}.
  \bibinfo{pages}{659--668}.
\newblock


\bibitem[Moffat and Zobel(2008)]%
        {moffat2008rank}
\bibfield{author}{\bibinfo{person}{Alistair Moffat} {and}
  \bibinfo{person}{Justin Zobel}.} \bibinfo{year}{2008}\natexlab{}.
\newblock \showarticletitle{Rank-biased precision for measurement of retrieval
  effectiveness}.
\newblock \bibinfo{journal}{\emph{ACM Transactions on Information Systems
  (TOIS)}} \bibinfo{volume}{27}, \bibinfo{number}{1} (\bibinfo{year}{2008}),
  \bibinfo{pages}{1--27}.
\newblock


\bibitem[Montazeralghaem et~al\mbox{.}(2016)]%
        {montazeralghaem2016axiomatic}
\bibfield{author}{\bibinfo{person}{Ali Montazeralghaem}, \bibinfo{person}{Hamed
  Zamani}, {and} \bibinfo{person}{Azadeh Shakery}.}
  \bibinfo{year}{2016}\natexlab{}.
\newblock \showarticletitle{Axiomatic analysis for improving the log-logistic
  feedback model}. In \bibinfo{booktitle}{\emph{Proceedings of the 39th
  International ACM SIGIR conference on Research and Development in Information
  Retrieval}}. \bibinfo{pages}{765--768}.
\newblock


\bibitem[Rahimi et~al\mbox{.}(2020)]%
        {rahimi2020axiomatic}
\bibfield{author}{\bibinfo{person}{Razieh Rahimi}, \bibinfo{person}{Ali
  Montazeralghaem}, {and} \bibinfo{person}{Azadeh Shakery}.}
  \bibinfo{year}{2020}\natexlab{}.
\newblock \showarticletitle{An axiomatic approach to corpus-based
  cross-language information retrieval}.
\newblock \bibinfo{journal}{\emph{Information Retrieval Journal}}
  \bibinfo{volume}{23}, \bibinfo{number}{3} (\bibinfo{year}{2020}),
  \bibinfo{pages}{191--215}.
\newblock


\bibitem[Roberts(1985)]%
        {roberts1985measurement}
\bibfield{author}{\bibinfo{person}{Fred~S Roberts}.}
  \bibinfo{year}{1985}\natexlab{}.
\newblock \showarticletitle{Measurement theory}.
\newblock \bibinfo{journal}{\emph{Encyclopedia of Mathematics and its
  applications}}  \bibinfo{volume}{7} (\bibinfo{year}{1985}).
\newblock


\bibitem[Rosenberg and Hirschberg(2007)]%
        {rosenberg2007v}
\bibfield{author}{\bibinfo{person}{Andrew Rosenberg} {and}
  \bibinfo{person}{Julia Hirschberg}.} \bibinfo{year}{2007}\natexlab{}.
\newblock \showarticletitle{V-measure: A conditional entropy-based external
  cluster evaluation measure}. In \bibinfo{booktitle}{\emph{Proceedings of the
  2007 joint conference on empirical methods in natural language processing and
  computational natural language learning (EMNLP-CoNLL)}}.
  \bibinfo{pages}{410--420}.
\newblock


\bibitem[Rosset et~al\mbox{.}(2019)]%
        {rosset2019axiomatic}
\bibfield{author}{\bibinfo{person}{Corby Rosset}, \bibinfo{person}{Bhaskar
  Mitra}, \bibinfo{person}{Chenyan Xiong}, \bibinfo{person}{Nick Craswell},
  \bibinfo{person}{Xia Song}, {and} \bibinfo{person}{Saurabh Tiwary}.}
  \bibinfo{year}{2019}\natexlab{}.
\newblock \showarticletitle{An axiomatic approach to regularizing neural
  ranking models}. In \bibinfo{booktitle}{\emph{Proceedings of the 42nd
  international ACM SIGIR conference on research and development in information
  retrieval}}. \bibinfo{pages}{981--984}.
\newblock


\bibitem[Rota(1971)]%
        {rota1971combinatorics}
\bibfield{author}{\bibinfo{person}{Gian-Carlo Rota}.}
  \bibinfo{year}{1971}\natexlab{}.
\newblock \showarticletitle{On the combinatorics of the Euler characteristic}.
  In \bibinfo{booktitle}{\emph{Studies in Pure Mathematics (Presented to
  Richard Rado)}}. Academic Press London, \bibinfo{pages}{221--233}.
\newblock


\bibitem[Sakai(2021)]%
        {sakai2021fuhr}
\bibfield{author}{\bibinfo{person}{Tetsuya Sakai}.}
  \bibinfo{year}{2021}\natexlab{}.
\newblock \showarticletitle{On Fuhr's guideline for IR evaluation}. In
  \bibinfo{booktitle}{\emph{ACM SIGIR Forum}}, Vol.~\bibinfo{volume}{54}. ACM
  New York, NY, USA, \bibinfo{pages}{1--8}.
\newblock


\bibitem[Sakai and Kando(2008)]%
        {sakai2008information}
\bibfield{author}{\bibinfo{person}{Tetsuya Sakai} {and} \bibinfo{person}{Noriko
  Kando}.} \bibinfo{year}{2008}\natexlab{}.
\newblock \showarticletitle{On information retrieval metrics designed for
  evaluation with incomplete relevance assessments}.
\newblock \bibinfo{journal}{\emph{Information Retrieval}} \bibinfo{volume}{11},
  \bibinfo{number}{5} (\bibinfo{year}{2008}), \bibinfo{pages}{447--470}.
\newblock


\bibitem[Sebastiani(2015)]%
        {sebastiani2015axiomatically}
\bibfield{author}{\bibinfo{person}{Fabrizio Sebastiani}.}
  \bibinfo{year}{2015}\natexlab{}.
\newblock \showarticletitle{An axiomatically derived measure for the evaluation
  of classification algorithms}. In \bibinfo{booktitle}{\emph{Proceedings of
  the 2015 international conference on the theory of information retrieval}}.
  \bibinfo{pages}{11--20}.
\newblock


\bibitem[Sebastiani(2020)]%
        {sebastiani2020evaluation}
\bibfield{author}{\bibinfo{person}{Fabrizio Sebastiani}.}
  \bibinfo{year}{2020}\natexlab{}.
\newblock \showarticletitle{Evaluation measures for quantification: An
  axiomatic approach}.
\newblock \bibinfo{journal}{\emph{Information Retrieval Journal}}
  \bibinfo{volume}{23}, \bibinfo{number}{3} (\bibinfo{year}{2020}),
  \bibinfo{pages}{255--288}.
\newblock


\bibitem[Sokolova(2006)]%
        {sokolova2006assessing}
\bibfield{author}{\bibinfo{person}{Marina Sokolova}.}
  \bibinfo{year}{2006}\natexlab{}.
\newblock \showarticletitle{Assessing invariance properties of evaluation
  measures}. In \bibinfo{booktitle}{\emph{Proceedings of the Workshop on
  Testing of Deployable Learning and Decision Systems, the 19th Neural
  Information Processing Systems Conference (NIPS 2006)}}.
\newblock


\bibitem[Swets(1963)]%
        {swets1963information}
\bibfield{author}{\bibinfo{person}{John~A Swets}.}
  \bibinfo{year}{1963}\natexlab{}.
\newblock \showarticletitle{Information retrieval systems}.
\newblock \bibinfo{journal}{\emph{Science}} \bibinfo{volume}{141},
  \bibinfo{number}{3577} (\bibinfo{year}{1963}), \bibinfo{pages}{245--250}.
\newblock


\bibitem[Van~Rijsbergen(1974)]%
        {van1974foundation}
\bibfield{author}{\bibinfo{person}{Cornelis~Joost Van~Rijsbergen}.}
  \bibinfo{year}{1974}\natexlab{}.
\newblock \showarticletitle{Foundation of evaluation}.
\newblock \bibinfo{journal}{\emph{Journal of documentation}}
  \bibinfo{volume}{30}, \bibinfo{number}{4} (\bibinfo{year}{1974}),
  \bibinfo{pages}{365--373}.
\newblock


\bibitem[V{\"o}lske et~al\mbox{.}(2021)]%
        {volske2021towards}
\bibfield{author}{\bibinfo{person}{Michael V{\"o}lske},
  \bibinfo{person}{Alexander Bondarenko}, \bibinfo{person}{Maik Fr{\"o}be},
  \bibinfo{person}{Benno Stein}, \bibinfo{person}{Jaspreet Singh},
  \bibinfo{person}{Matthias Hagen}, {and} \bibinfo{person}{Avishek Anand}.}
  \bibinfo{year}{2021}\natexlab{}.
\newblock \showarticletitle{Towards Axiomatic Explanations for Neural Ranking
  Models}. In \bibinfo{booktitle}{\emph{Proceedings of the 2021 ACM SIGIR
  International Conference on Theory of Information Retrieval}}.
  \bibinfo{pages}{13--22}.
\newblock


\bibitem[Yao(1995)]%
        {yao1995measuring}
\bibfield{author}{\bibinfo{person}{YY Yao}.} \bibinfo{year}{1995}\natexlab{}.
\newblock \showarticletitle{Measuring retrieval effectiveness based on user
  preference of documents}.
\newblock \bibinfo{journal}{\emph{Journal of the American Society for
  Information science}} \bibinfo{volume}{46}, \bibinfo{number}{2}
  (\bibinfo{year}{1995}), \bibinfo{pages}{133--145}.
\newblock


\bibitem[Zhai and Fang(2013)]%
        {zhai2013axiomatic}
\bibfield{author}{\bibinfo{person}{ChengXiang Zhai} {and} \bibinfo{person}{Hui
  Fang}.} \bibinfo{year}{2013}\natexlab{}.
\newblock \showarticletitle{Axiomatic analysis and optimization of information
  retrieval models}. In \bibinfo{booktitle}{\emph{Proceedings of the 2013
  Conference on the Theory of Information Retrieval}}. \bibinfo{pages}{3--3}.
\newblock


\end{thebibliography}

\appendix

\section{Formal Proofs}
\begin{proof}
{\small
{\bf [Proposition \ref{prop:operations-set}]:} 

It will be seen that $\big\{ \max \{ \hat{r}_1 , \hat{s}_1 \}$, $\ldots$, $\max \{ \hat{r}_N$, $\hat{s}_N \} \big\}$ is the lower upper bound of $\hat{r}$ and $\hat{s}$.

First, it is an upper bound of $\hat{r}$ and $\hat{s}$ since $\hat{r}_i \preccurlyeq \max \{ \hat{r}_i , \hat{s}_i \}$, $\hat{s}_i \preccurlyeq \max \{ \hat{r}_i , \hat{s}_i \}$, $\forall i \in \{1, \ldots, N\}$, and these conditions fulfil the definition of the \textbf{replacement} [set-based] partial ordering.

Now, it will be seen that it is the lower upper bound. Let $\hat{t} = \{\hat{t}_1, \ldots, \hat{t}_N\} \in R(N)$, such that $\hat{r} \preceq \hat{t}$ and $\hat{s} \preceq \hat{t}$. Then, $\big\vert\{i : \hat{r}_i \succcurlyeq \mathit{a}_j\}\big\vert \leq \big\vert\{i : \hat{t}_i \succcurlyeq \mathit{a}_j\}\big\vert, \forall j \in \{0, \ldots, c\}$ and $\big\vert\{i : \hat{s}_i \succcurlyeq \mathit{a}_j\}\big\vert \leq \big\vert\{i : \hat{t}_i \succcurlyeq \mathit{a}_j\}\big\vert, \forall j \in \{0, \ldots, c\}$. Therefore, $\big\vert\{i : \max(\hat{r}_i,\hat{s}_i) \succcurlyeq \mathit{a}_j\}\big\vert \leq \big\vert\{i : \hat{t}_i \succcurlyeq \mathit{a}_j\}\big\vert, \forall j \in \{0, \ldots, c\}$.

The other equality can be seen by duality.
}
\end{proof}

\begin{proof}
{\small
{\bf [Proposition \ref{prop:distrib-set}]:}

The \textbf{replacement} [set-based] partial order is grounded on the \emph{replacement} operation. It will be seen that one system run, $\hat{s} \in R(N)$, covers another, $\hat{r} \in R(N)$, if they differ only in the \emph{replacement} of a document with the relevance degree immediately consecutive in $REL$. Formally:
$$\hat{r} \precdot \hat{s} \Longleftrightarrow \exists k : \hat{r}_k = \mathit{a}_i, \hat{s}_k = \mathit{a}_{i+1} \wedge \hat{r}_j = \hat{s}_j, \forall j \neq k$$

This is true, since if $\exists k : \hat{r}_k = \mathit{a}_i, \hat{s}_k = \mathit{a}_{i+1} \wedge \hat{r}_j = \hat{s}_j, \forall j \neq k$, then $\hat{s}$ has more \emph{cumulative mass of relevance} than $\hat{r}$ for every relevance degree, which implies that $\hat{r} \preceq \hat{s}$. In addition, there cannot be a different judged run between $\hat{r}$ and $\hat{s}$ since they differ in only a document that has been classified with the consecutive relevance degree.

The aim is to apply Theorem \ref{th:distributive_n_m} to conclude that $(R(N), \wedge, \vee, \preceq)$ is a distributive lattice. Thus, it is necessary to see that $R(N)$ does not contain a sublattice equal to $N_5$ or $M_3$.

Suppose that $R(N)$ contains a sublattice $N_5$ as depicted in Figure \ref{fig:distributive}, where a = $\hat{u}$, b = $\hat{r}^1$, c = $\hat{r}^2$, d = $\hat{s}$ and e = $\hat{v}$. It can be postulated that $\hat{u} = \{\mathit{a}_i, \mathit{a}_j, \mathit{a}_k\}$, where $\mathit{a}_i, \mathit{a}_j, \mathit{a}_k \in REL$. Perhaps, $\hat{u}$ may contain more relevance degrees; however, they do not participate in the demonstration. Thus, this hypothesis can be assumed to hold.

According to the covering relation, $\hat{r}^1$ and $\hat{s}$ have to increase one relevance degree of $\hat{u}$, being consecutives in $REL$. Notice that they cannot increase the same relevance degree, since they are placed in separate paths, in the Hasse diagram. Thus, it can be assumed that $\hat{r}^1 = \{\mathit{a}_{i+1}, \mathit{a}_j, \mathit{a}_k\}$ and $\hat{s} = \{\mathit{a}_i, \mathit{a}_{j+1}, \mathit{a}_k\}$.

Then, $\hat{r}^2$ has to increase one of the relevance degrees of $\hat{r}^1$. There are three possibilities, which will be analysed in the following:

\begin{enumerate}
\item Suppose that $\mathit{a}_{i+1}$, has been increased, i.e., $\hat{r}^2 = \{\mathit{a}_{i+2}, \mathit{a}_j, \mathit{a}_k\}$. Then, the configuration of $\hat{v}$ would contain $\mathit{a}_{i+2}$ or $\mathit{a}_{i+3}$, depending on if there is an increase in that relevance degree. However, both cases are not possible when the covering relation is applied to $\hat{s}$ since $\hat{s} = \{\mathit{a}_i, \mathit{a}_{j+1}, \mathit{a}_k\}$, and $\mathit{a}_i$ can increase only one unity in $REL$.
\item Suppose that $\mathit{a}_{k}$ has been increased, i.e., $\hat{r}^2 = \{\mathit{a}_{i+1}, \mathit{a}_j, \mathit{a}_{k+1}\}$. Then, this configuration is not compatible with $\hat{s}$ since it can increase only $\mathit{a}_{i}$ or $\mathit{a}_{k}$ but not both.
\item Suppose that $\mathit{a}_j$ has been increased, i.e., $\hat{r}^2 = \{\mathit{a}_{i+1}, \mathit{a}_{j+1}, \mathit{a}_{k}\}$. Then, $\hat{s}$ has to increase $\mathit{a}_{i}$ and any increase in $\hat{r}^2$ will not be compatible with $\hat{s}$.
\end{enumerate}

As every case is not possible, $R(N)$ does not contain any $N_5$. 

Now, suppose that $R(N)$ contains a sublattice $M_3$ as depicted in Figure \ref{fig:distributive}, where a = $\hat{u}$, b = $\hat{r}$, c = $\hat{s}$, d = $\hat{t}$ and e = $\hat{v}$. It can be postulated that $\hat{u} = \{\mathit{a}_i, \mathit{a}_j, \mathit{a}_k\}$, where $\mathit{a}_i, \mathit{a}_j, \mathit{a}_k \in REL$. Perhaps, $\hat{u}$ may contain more relevance degrees; however, they do not take part in the demonstration. Thus, this hypothesis can be assumed to hold.

According to the covering relation, $\hat{r}$, $\hat{s}$ and $\hat{t}$ have to increase one relevance degree of $\hat{u}$, being consecutive in $REL$. Note that they cannot increase the same relevance degree since they are placed in separate paths in the Hasse diagram. Thus, it can be assumed that $\hat{r} = \{\mathit{a}_{i+1}, \mathit{a}_j, \mathit{a}_k\}$, $\hat{s} = \{\mathit{a}_i, \mathit{a}_{j+1}, \mathit{a}_k\}$ and $\hat{t} = \{\mathit{a}_i, \mathit{a}_{j}, \mathit{a}_{k+1}\}$.

Then, $\hat{v}$ will be obtained by increasing one relevance degree of $\hat{r}$, $\hat{s}$ and $\hat{t}$. For example, if $\hat{r}$ increases $\mathit{a}_j$, then $\hat{v} = \{\mathit{a}_{i+1}, \mathit{a}_{j+1}, \mathit{a}_k\}$. This configuration is compatible with $\hat{s}$, when $\mathit{a}_i$ is increased. However, $\hat{v}$ is not compatible with $\hat{t}$ since it can only increase $\mathit{a}_i$ or $\mathit{a}_j$ but not both. Similar reasoning shows that the rest of the cases are not possible.

Thus, $R(N)$ cannot contain $M_3$ as a sublattice. Then, according to Theorem \ref{th:distributive_n_m}, $R(N)$ is a distributive lattice.
}
\end{proof}

\begin{proof}
{\small
{\bf [Proposition \ref{prop:irred-set}]:}

In the demonstration of Proposition \ref{prop:distrib-set}, it is shown that the covering relation of the \textbf{replacement} [set-based] ordering is as follows: given a pair of system runs, $\hat{r}, \hat{s} \in R(N)$ with $\hat{r} \neq \hat{s}$, then $\hat{r} \precdot \hat{s}$ $\Leftrightarrow$ $\exists k : \hat{r}_k = \mathit{a}_i, \hat{s}_k = \mathit{a}_{i+1}$ and $\hat{r}_j = \hat{s}_j, \forall j \neq k$.

The join-irreducible elements are those which only have one descendant in the Hasse diagram. It will be seen that these elements are those which only have two different relevance degrees in the configuration, and one of them is $\mathit{a}_0$.

Let $\hat{r}$ be a judged run with at least two documents classified with relevance degrees different from $\mathit{a}_0$. Thus, $\hat{r} = \{\mathit{a}_i, \mathit{a}_j\} \cup \hat{r'}$, where $i$, $j \neq 0$ and $\hat{r'}$ could be any subset of judged documents, even the empty set. Then, according to the covering relation, $\hat{r}$ has at least two descendants in the Hasse diagram. They are: $\{\mathit{a}_{i-1}, \mathit{a}_j,\} \cup \hat{r'}$ and $\{\mathit{a}_i, \mathit{a}_{j-1}\} \cup \hat{r'}$, thus $\hat{r}$ cannot be a join-irreducible element.

Finally, the judged runs listed in the statement of this Proposition are the only ones which have one or two relevance degrees, where one of them is $\mathit{a}_0$. It will be seen that these elements only have one descendant in the Hasse diagram. Let $\hat{r} = \{\mathit{a}_i, \ldots, \mathit{a}_i, \mathit{a}_0, \ldots, \mathit{a}_0\}$ be one of these elements, then the only descendant in the Hasse diagram is $\{\mathit{a}_i, \ldots, \mathit{a}_i, \mathit{a}_{i-1}, \mathit{a}_0, \ldots, \mathit{a}_0\}$ since it only can have one relevance degree different and consecutive in $REL$.
}
\end{proof}

\begin{proof}
{\small
{\bf [Proposition \ref{prop:valuation-set}]:}

Given two system runs $\hat{r}$, $\hat{s} \in R(N)$, it will be seen that $gP(\hat{r} \vee \hat{s}) + gP(\hat{r} \wedge \hat{s}) = gP(\hat{r}) + gP(\hat{s})$. Taking into account the definition of the meet and the join in Proposition \ref{prop:operations-set},
$$gP(\hat{r} \vee \hat{s}) + gP(\hat{r} \wedge \hat{s}) = \frac{1}{N} \cdot \sum_{i=1}^{N} \frac{\max\{g(\hat{r}_i), g(\hat{s}_i)\}}{g(\mathit{a}_{c})} + \frac{1}{N} \cdot \sum_{i=1}^{N} \frac{\min\{g(\hat{r}_i), g(\hat{s}_i)\}}{g(\mathit{a}_{c})} = $$
Considering the noteworthy property of real numbers: $\max\{x,y\}+ \min\{x,y\}=x+y$, it holds:
\begin{align*}
=\frac{1}{N} \cdot \sum_{i=1}^{N} \frac{\max\{g(\hat{r}_i), g(\hat{s}_i)\} + \min\{g(\hat{r}_i), g(\hat{s}_i)\}}{g(\mathit{a}_{c})} & = \frac{1}{N} \cdot \sum_{i=1}^{N} \frac{g(\hat{r}_i) + g(\hat{s}_i)}{g(\mathit{a}_{c})}= \\
=\frac{1}{N} \cdot \sum_{i=1}^{N} \frac{g(\hat{r}_i)}{g(\mathit{a}_{c})} + \frac{1}{N} \cdot \sum_{i=1}^{N} \frac{g(\hat{s}_i)}{g(\mathit{a}_{c})} & =gP(\hat{r}) + gP(\hat{s})
\end{align*}
The same demonstration is valid for the generalized recall, where $N$ should changed by $RB$.
}
\end{proof}

\begin{proof}
{\small
{\bf [Proposition \ref{prop:operations-rank}]:}
The demonstration is quite similar to the proof of Proposition \ref{prop:operations-set}.
It will be seen that $\left( \max \{ \hat{r}_1 , \hat{s}_1 \}, \ldots, \max \{ \hat{r}_N , \hat{s}_N \} \right)$ is the lower upper bound of $\hat{r}$ and $\hat{s}$.

First, it is an upper bound of $\hat{r}$ and $\hat{s}$ since $\hat{r}[i] \preccurlyeq \max \{ \hat{r}[i] , \hat{s}[i] \}$, $\hat{s}[i] \preccurlyeq \max \{ \hat{r}[i] , \hat{s}[i] \}$, $\forall i \in \{1, \ldots, N\}$, and these conditions fulfil the definition of the \textbf{replacement} [rank-based] partial ordering.

Now, it will be seen that it is the lower upper bound. Let $\hat{t} = (\hat{t}[1], \ldots, \hat{t}[N])$ $\in R(N)$, such that $\hat{r} \preceq \hat{t}$ and $\hat{s} \preceq \hat{t}$. Then, $\big\vert\{i : \hat{r}[i] \succcurlyeq \mathit{a}_j\}\big\vert \leq \big\vert\{i : \hat{t}[i] \succcurlyeq \mathit{a}_j\}\big\vert, \forall j \in \{0, \ldots, c\}$ and $\big\vert\{i : \hat{s}[i] \succcurlyeq \mathit{a}_j\}\big\vert \leq \big\vert\{i : \hat{t}[i] \succcurlyeq \mathit{a}_j\}\big\vert, \forall j \in \{0, \ldots, c\}$. Therefore $\big\vert\{i : \max(\hat{r}[i],\hat{s}[i]) \succcurlyeq \mathit{a}_j\}\big\vert \leq \big\vert\{i : \hat{t}[i] \succcurlyeq \mathit{a}_j\}\big\vert, \forall j \in \{0, \ldots, c\}$.

The other equality can be seen by duality.
}
\end{proof}

\begin{proof}
{\small
{\bf [Proposition \ref{prop:irred-rank}]:}
The demonstration is quite similar to the proof of Proposition \ref{prop:irred-set}.

In the demonstration of the Proposition \ref{prop:distrib-rank}, it is shown that the covering relation of the \textbf{replacement} [rank-based] ordering is as follows: given a pair of system runs, $\hat{r}, \hat{s} \in R(N)$ with $\hat{r} \neq \hat{s}$, then $\hat{r} \precdot \hat{s} \Leftrightarrow \exists k : \hat{r}[k] = \mathit{a}_i, \hat{s}[k] = \mathit{a}_{i+1}$ and $\hat{r}[j] = \hat{s}[j], \forall j \neq k$.

The join-irreducible elements are those which only have one descendant in the Hasse diagram. It will be seen that these elements are those which only have one component different from $\mathit{a}_0$.

Let $\hat{r}$ be a judged run with at least two components with relevance degrees different from $\mathit{a}_0$. Thus, $\hat{r} = (\ldots,\mathit{a}_i, \ldots, \mathit{a}_j, \ldots)$, where $i$, $j \neq 0$. Then, according to the covering relation, $\hat{r}$ has at least two descendants in the Hasse diagram. They are: $(\ldots, \mathit{a}_{i-1}, \ldots, \mathit{a}_j, \ldots)$ and $(\ldots, \mathit{a}_i,$ $\ldots, \mathit{a}_{j-1}, \ldots)$, thus $\hat{r}$ cannot be a join-irreducible element.

Finally, the judged runs listed in the statement of this Proposition are the ones which only have one component different from $\mathit{a}_0$. It will be seen that these elements only have one descendant in the Hasse diagram. Let $\hat{r} = (\mathit{a}_0, \ldots, \mathit{a}_i, \ldots, \mathit{a}_0)$ be one of these elements, then the only descendant in the Hasse diagram is $(\mathit{a}_0, \ldots, \mathit{a}_{i-1}, \ldots, \mathit{a}_0)$ since it only can have one relevance degree different and consecutive in $REL$.
}
\end{proof}

{\small
The proofs of Propositions %\ref{prop:operations-rank}, 
\ref{prop:distrib-rank} %, \ref{prop:irred-rank} 
and \ref{prop:valuation-rank} 
are quite similar to the proofs of Propositions %\ref{prop:operations-set}, 
\ref{prop:distrib-set} %, \ref{prop:irred-set} and 
and \ref{prop:valuation-set}, respectively, which have been omitted for space limitations.
}

\end{document}